**Institute of Metallurgy and Materials Science, Polish Academy of Sciences**

**Institute of Physics, Jagiellonian University**

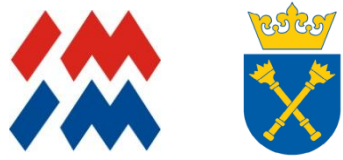

# Development of novel plastic scintillators based on polyvinyltoluene for the hybrid J-PET/MR tomograph

Anna Wieczorek

**Doctoral Dissertation**

prepared in the Institute of Physics, Jagiellonian University

Supervisor: prof. dr hab. Paweł Moskal

Institute of Physics, Jagiellonian University

Co-supervisor: dr Andrzej Kochanowski

Faculty of Chemistry, Jagiellonian University

**Kraków, 2017**

# Abstract

A novel plastic scintillator, referred to as J-PET scintillator, has been developed for the application in the digital positron emission tomography (PET). The novelty of the concept lies in application of the 2-(4-styrylphenyl)benzoxazole as a wavelength shifter. To date, the chemical compound has not been used as scintillator additive. A role of wavelength shifter is to shift the scintillation spectrum towards longer wavelengths making it more suitable for applications in scintillators of long strips geometry because of larger attenuation length, and light detection with digital silicon photomultipliers.

In the thesis method of scintillators development and characterization of their properties are presented as well as synthesis method of the novel wavelength shifter. The optimal concentration of the novel wavelength shifter was established by maximizing the light output. Performance of scintillator, considering light emission spectrum, light emission efficiency, rising and decay times of the scintillation pulses were analyzed and compared to state of the art commercially available plastic scintillators. Structure of J-PET scintillators was studied and discussed.

# Streszczenie

W niniejszej pracy została przedstawiona charakterystyka nowatorskiego scyntylatora J-PET. Docelowo scyntylator zastosowany będzie w cyfrowej pozytonowej tomografii emisyjnej (PET). Innowacyjność scyntylatora polega na zastosowaniu 2-(4-styrylofenylo)benzoksazolu jako przesuwacza długości fali. Do tej pory ten związek chemiczny nie był stosowany jako dodatek scyntylacyjny. Jego zadaniem jest przesunięcie widma emisji w kierunku dłuższych fal. Dzięki temu możliwa jest efektywna detekcja światła scyntylacyjnego przez fotopowielacze krzemowe. Ponadto, przesunięcie widma emisji w kierunku dłuższych fal pozwala na zastosowanie 2-(4-styrylofenylo)benzoksazolu w długich paskach scyntylatorów ze względu na większą drogę tłumienia.

W niniejszej pracy została przedstawiona metoda wytwarzania nowatorskich scyntylatorów polimerowych oraz ich właściwości, a także metoda syntezy nowego przesuwacza długości fali. Zostało zoptymalizowane stężenie 2-(4-styrylofenylo)benzoksazolu umożliwiające osiągnięcie maksymalnej wydajności świetlnej. Działanie scyntylatora zostało scharakteryzowane biorąc pod uwagę widmo emisji, wydajność świetlną oraz czasy narastania i zaniku impulsów świetlnych. Wartości poszczególnych parametrów zostały porównane z odpowiednimi parametrami najlepszych scyntylatorów komercyjnych. Przedmiotem badań opisanych w pracy jest również struktura scyntylatorów J-PET.

# Contents

# 1. Introduction

Positron Emission Tomography (PET) has been a research tool for about fifty years. Nowadays, it is well known mainly from the application to clinical medicine. This technique enables quantitative, three-dimensional images for the study of physiological and biochemical processes occurring in the human body.

PET scanners are valuable diagnostic devices because of the ability of cancer detection even in its early stages, what allows to take an immediate treatment increasing the chances of the patient for the recovery. That also enables adjustment of the proper therapy and judge of its effectiveness. PET is used for diagnosis of many other diseases, like Alzheimer, Parkinson, cardiology, neurological and gastrological diseases.

At the Jagiellonian University, a novel PET scanner has been designed and built. The device is called J-PET (Jagiellonian PET). The main difference between J-PET and conventional PET scanners is utilization of long plastic scintillators strips instead of small inorganic crystals. Scintillators are the key part of tomography scanners. They enable detection of ionizing radiation which is emitted from the patient during the examination. In the PET technique a patient is administered radiopharmaceuticals emitting positrons. During annihilation of incident positrons with electrons from human body, gamma quanta are created. They interact with the scintillating material and as a result light pulses are produced. The light is subsequently detected by photomultipliers connected to scintillators. Plastic scintillators utilized in J-PET device decreases costs of the scanner significantly and open perspectives for examination of a large part of human body during one scan.

J-PET technology also opens perspective for the simultaneous PET and Magnetic Resonance Imaging (PET/MRI) and for the construction of PET inserts which may be adapted to nowadays MR scanners held by hospitals. Such hybrid device, J-PET/MR scanner will enable simultaneous anatomical and functional imaging. However several changes should be made in the first J-PET prototype to be MR compatible, e.g. it is essential to exchange traditional vacuum photomultipliers to digital silicon photodetectors.

Proper scintillators need to be adjusted to the hybrid J-PET/MR scanner. They should be characterized by several parameters: high light output, long attenuation length and their emission spectra should fit quantum efficiency of silicon photomultipliers. Light output characterizes the light emission efficiency of scintillators. Long attenuation length of light in the scintillating material is important especially for J-PET/MRI because of the requirement of application of long scintillator strips, through which light needs to be transferred effectively.

The aim of the dissertation was development of novel plastic scintillators to J-PET/MR scanner and characterization of their properties.

The thesis which will be proved is as follows: in laboratory condition it is possible to obtain plastic scintillator characterized by high light output, weak light absorption in the material and with emission spectrum matched to quantum efficiency of silicon photomultipliers.

This thesis concentrates on development of novel plastic scintillator, referred to as J-PET scintillator, and characterization of its properties. The manuscript is organized in the following way:

- Chapter 2 outlines theoretical motivation for conducted investigations and introduces novel J-PET concept.
- Chapter 3 comprises explanation of scintillation mechanism and includes description of state of the art plastic scintillator offered by worldwide companies. Current application of novel scintillating dopant is described as well.
- In Chapter 4, experimental methods are shortly described.
- Further on, in Chapter 5, three chemical compounds, tested as wavelength shifter, as well as their synthesis schemes are described. Plastic scintillators containing novel dopants were prepared and preliminary results of their performance are presented. Scintillators with 2-(4-styrylphenyl)benzoxazole, acted at most efficiently and were a subject of research presented in next chapters.
- Chapter 6 includes detailed information about bulk polymerization process, which is used for development of J-PET plastic scintillators.
- In Chapter 7, optical, spectral and timing properties of novel plastic scintillators are described. Optical adjustment of particular components of scintillators, regarding their emission and absorption spectra are discussed. The most important parameter characterizing scintillators performance: the light output was determined and evaluated with respect to commercially available scintillators. Timing properties of J-PET scintillators and comparison to commercial scintillators in view of application in the J-PET system are shown.

- In Chapter 8, structure of J-PET scintillators were analyzed. Molecular weight was estimated, proving assurance of maximal light output considering polymer structure impact. Further on, investigations of scintillators structure with Positron Annihilation Lifetime Spectroscopy (PALS) and Differential Scanning Calorimetry (DSC) were presented and compared.
- Chapter 9 comprises description of development of plastic scintillator strips with large dimensions.
- In Chapter 10, summary and final conclusions followed by perspectives of further researches are presented.

## 2. Positron Emission Tomography

In PET scanners used nowadays in hospitals the ring of inorganic crystal scintillators plays a role of detector. In the examination with PET, the patient is administered pharmaceutical marker (e.g. 18-fluorodeoxyglucose - FDG) labeled with the radioactive isotope decaying by the emission of positron ($\beta^+$) [1] [2]. The positron annihilates with an electron from the body of patient. As a result of annihilation, the mass of the positron and electron is converted predominantly into two gamma quanta with energy 511 keV each, emitted in opposite directions. This phenomenon is a basis of the PET image reconstruction. Hit positions of both gamma quanta on scintillator ring is determined and the line of response (LOR) can be marked. Intersection of many lines of response allows to determine the annihilation point. A scheme of annihilation and PET scanner is shown in Fig 1.

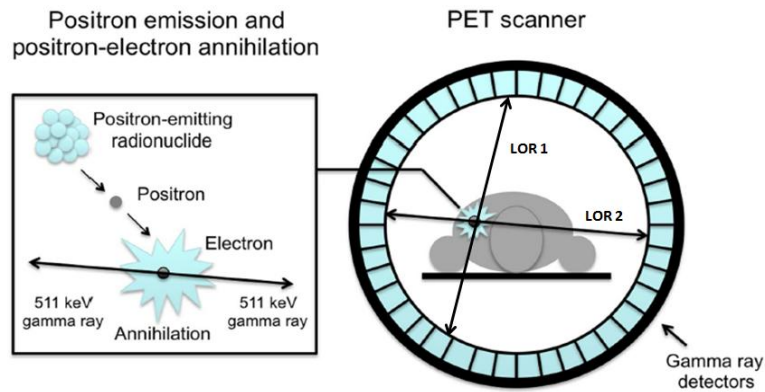

**Figure 1 The scheme of positron - electron annihilation (left) and PET scanner (right). Figure is adapted from [3].**

From the PET image, the tracer distribution in the human body can be quantitatively determined. The diagnostics is based on different rate of FDG consumption in different parts of the body. Cancer or diseased cells have larger metabolism of glucose in comparison to healthy tissues. This regions are clearly visible on the PET image [1] [3] [4] [5].

Scanners of positron emission tomography based on inorganic crystals are very expensive, mainly because of high price of the scintillators. In Tab. 1 prices per 1 $cm^3$ of three inorganic scintillators that may be used in nowadays PET scanners are given.

**Table 1 Prices of inorganic scintillators possible to use in PET [6].**

| Scintillator | Price per 1 $cm^3$ |
|---|---|
| BGO | $35 |
| LSO:Ce | $60 |
| LYSO:Ce | $70 |

Current diagnostic chambers of PET scanners consist of a ring containing a large amount of inorganic scintillators in the form of small size rectangular blocks. PET scanners e.g. produced by GE Healthcare, are built of about 12,000 of BGO inorganic crystals, 0.9 $cm^3$ each. So there is a need of spending even more than $350,000 on scintillators, not including electronics or other elements of the tomography scanner. That makes the price of conventional PET scanners, and what is more, the price of single tomography scan extremely high [1]. Thus diagnostics using nowadays positron emission tomography is beyond the reach of many countries in the world.

## 2.1. Jagiellonian PET: J-PET

Due to the growing demand for Positron Emission Tomography scanners, which in Poland and other countries is still insufficient, an innovative method for the construction of these devices was developed [7] [8] [9] [10] [11]. Expensive inorganic scintillators were replaced by low-cost plastic scintillators. 1 cm$^3$ of plastic scintillator costs approximately $1. Usage of plastic scintillators instead of inorganic ones decreases price of the whole scanner significantly.

So far plastic scintillators have not been investigated as detectors of gamma quanta in PET scanners because of much lower detection efficiency comparing to inorganic crystals and negligible probability of photoelectric effect. Gamma quanta interact with low atomic number elements, which plastic scintillators are made of, predominantly via Compton effect. That leads to continuous charge spectra. In addition, efficiency of gamma quanta detection is much lower than in inorganic crystals. This disadvantage can be compensated by very good timing properties of plastic scinitillators and by the large size of the diagnostic chamber.

In Table 2 selected parameters describing properties of BC (Saint Gobain) and EJ (Eljen Technology) plastic scintillators and four inorganic crystals (BGO, GSO, LSO and LYSO) that are used in traditional PET scanner are given. Light outputs of LSO or LYSO crystals are much higher in comparison to plastic scintillators. However, attenuation length in plastic scintillator is longer than in crystals. Because plastic scintillators can be produced in any shapes and sizes, longer attenuation length makes possibility to design and manufacture long scintillator strips in which the light will be transferred effectively even at large distances. This in turn enables to construct PET scanner basing on long scintillator strips with large field of view. In such scanner imaging of whole patient body or its large part will be possible.

**Table 2** Selected parameters of plastic scintillators offered by Saint Gobain (BC) and Eljen Technology (EJ) [12] [13] [14] [15] and inorganic crystal scintillators used in commercial PET solutions [16] [17].

| Scintillator | Light output [photons/MeV] | Light attenuation length [cm] | Decay time [ns] | Density [g/cm$^3$] |
|:---:|:---:|:---:|:---:|:---:|
| BC-400 | 10000 | 250 | 2.4 | 1.032 |
| BC-404 | 10880 | 160 | 1.8 | 1.032 |
| BC-408 | 10240 | 380 | 2.1 | 1.032 |
| BC-420 | 10240 | 110 | 1.5 | 1.032 |
| EJ-230 | 10240 | 120 | 1.5 | 1.023 |
| EJ-204 | 10880 | 160 | 1.8 | 1.023 |
| EJ-212 | 10000 | 250 | 2.4 | 1.023 |
| BGO | 6000 | 22.8 | 300 | 7.13 |
| GSO | 10000 | 22.2 | 50 | 6.71 |
| LSO | 29000 | 20.9 | 40 | 7.40 |
| LYSO | 18000 | 20.9 | 40 - 44 | 7.30 |

Decay time of light pulses in plastic scintillators are tens times shorter than in inorganic crystals. To take full advantage of fast signals, a novel front-end electronics [18], methods of image reconstruction [19] and data acquisition [20] were developed. The concept of low-price PET scanner and novel solutions related with its realization are the subject of 17 patents and patent applications by members of the J-PET Collaboration [7] [8].

Scheme of the simplest detection unit of J-PET system capable to register annihilation gamma quanta is shown in Fig. 2. To each plastic scintillator strip photomultipliers (PM) are connected at both ends. They play a role of converters of light into electric signals. Presently, Hamamatsu vacuum photomultipliers are utilized [21]. Times of the light signals arrivals to both photomultipliers of each pair (PM 1 - PM 2 and PM 3 - PM 4) are measured. From the difference between times of arrivals to the ends of strip, gamma quantum hit position can be determined and the point of annihilation along the line of response (LOR) can be calculated.

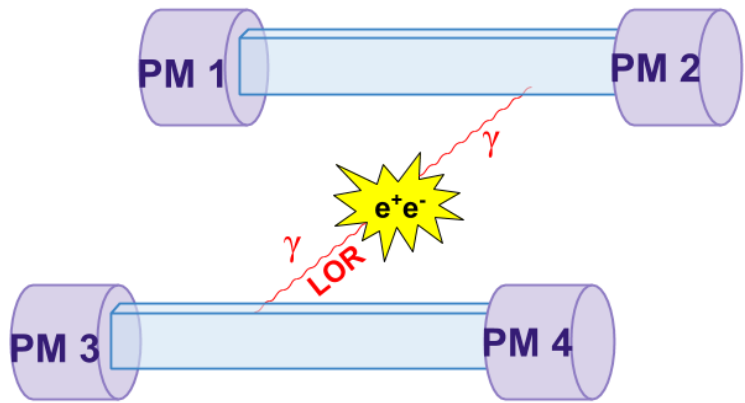

**Figure 2 Scheme of the simplest unit of J-PET scanner. PM denotes photomultiplier.**

First prototype of J-PET scanner has been already built. Presently it consists of 192 plastic scintillator strips, each 50 cm long. Thus the axial field of view of the J-PET prototype is equal to 50 cm and is few times larger compared to the presently available PET scanners based on crystal scintillators [22].

In J-PET scanner scintillators are wrapped into reflective and black foils, what improves the effectiveness of light detection and ensures light-tightness. The novel scanner has a potential of imaging large parts of human body simultaneously. Photographs of J-PET scanner prototype are shown in Fig. 3.

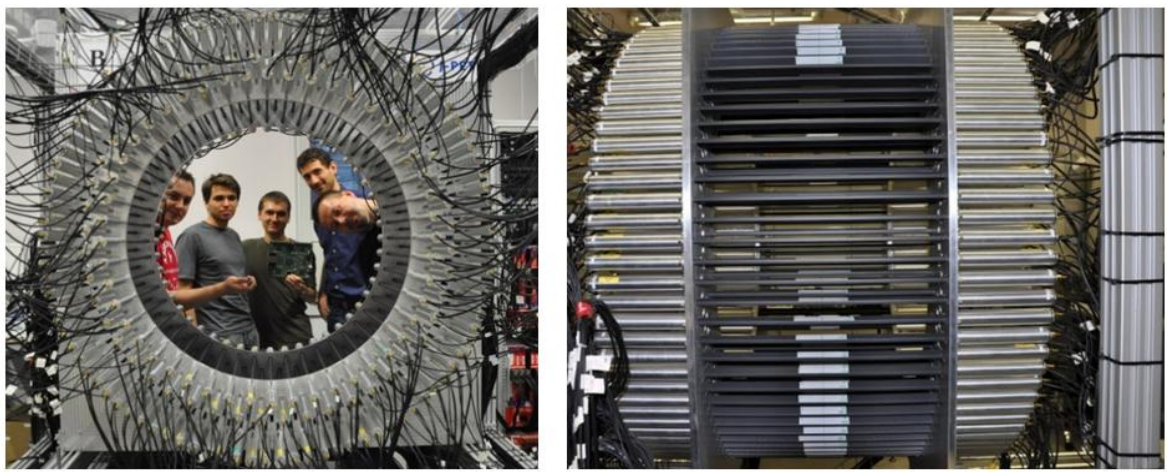

**Figure 3 A front (left) and side (right) view of J-PET scanner prototype** [23]**. Scintillators wrapped in the black foil are placed in the middle. At their both ends vacuum photomultipliers are connected. Photomultipliers are placed inside lightproof aluminum housings.**

## 2.2. Concept of novel J-PET/MR scanner

J-PET opens new possibilities in the field of diagnostics: allowing e.g. the simultaneous examination of the patient using two well established diagnostic methods: Positron Emission Tomography and Magnetic Resonance Imaging (MRI). Inventions based on this solution are subjects of patent applications [24] [25] [26].

Nowadays, hybrid PET/MR imaging devices are available, allowing simultaneous diagnostics with the two techniques. Such scanners are installed in two medical centers in Poland. However, the scanners are extremely expensive. Presently used scanners are based on inorganic crystals, like in conventional PET. J-PET/MR solution will be built of plastic scintillator, what will make the hybrid imaging technology more achievable.

Combining of both devices into a single one enables metabolic and morphological imaging during a single examination providing better diagnosis and treatment monitoring. This solution helps to eliminate artifacts in tomographic images which hinder identification of cancerous lesions. Typically it is possible to carry out exams with this two methods sequentially. Patient is positioned differently in PET and MRI scanners, therefore internal organs may move involuntary even if the time between both scans is short. In effect, alignment of images acquired during such exams is complicated.

J-PET/MR scanner is also a space and cost saving invention. Moreover, such hybrid device reduces radiation dose to which the patient will be exposed by providing both: morphological and functional imaging at the same time [27]. Morphological resolution achieved with MR is high, therefore combining with PET shows better contrast in soft tissues e.g. brain in comparison to hybrid PET/CT examination which is presently available [28].

The novel single hybrid device allowing simultaneous examination by PET and MRI methods is based on the J-PET concept. It consists of single detection modules built of plastic scintillators to which photomultipliers are connected at both ends. Instead of vacuum photomultipliers utilized in the first J-PET prototype, silicon ones are used.

Sillicon photomultipliers (SiPMs) are not affected by the magnetic field of MR scanner, equal to about 1.5 - 3 Tesla typically, and they do not disturb the homogeneity of magnetic field, so they can be utilized in the novel device.

J-PET device will be an insert possible to use in existing MR scanners held by hospitals without necessity of hardware modification. What is more, combining PET and MRI into one device will shorten the time of examination and will have positive impact on patient comfort [29].

## 3. Scintillators and scintillation mechanism

Ionising radiation interacting with matter excites its molecules. When they return to the ground state, photons of the visible or near to visible light spectrum range are produced. This phenomenon is called scintillation. A material in which conversion of excitation energy into light is highly efficient is called scintillator. According to [30] [31], scintillator used in the radiation detectors should be characterized by several properties:

- transparent at the wavelength of emitted scintillation light
- high efficiency of light production
- short light pulses to exclude delayed light emission
- the amount of light should be proportional to the energy deposited in the material
- chemical and mechanical stability
- not prone to radiation damage. It was estimated that one plastic scintillator with dimensions of 0.7 cm x 1.9 cm x 50 cm during one year of utilization in PET scanners would receive a radiation dose about 0.1 kGy. According to article [32], this is more than one order of magnitude less than the dose causing noticeable radiation damage in plastic scintillators.

Nowadays, scintillation detectors are one of the most popular detectors of radiation. Scintillators are common gamma ray, X-ray, charged and neutral particles detectors. They are utilized in many fields of science and industry. They are the most common detectors used in experiments, regarding the fundamental research in particle and nuclear physics, e.g. to detect particles formed during the process of artificial fusion of atomic nuclei. They are also used as detectors of cosmic rays and in the detectors systems installed at the Large Hadron Collider at the European Centre for Nuclear Research (CERN) in Geneva. Scintillators are widely used in astrophysics to observe emerging stars, the searches for mineral resources and to ensure security at the airports. One can use them on minefields to help in locating the explosive materials without endangering human life.

Scintillators are divided into two groups: inorganic and organic. Physics of scintillation mechanism as well as their properties and applications are different. The main difference is that organic scintillators predominantly consist of low atomic number (Z) elements, like carbon and hydrogen, and they have relatively long attenuation length. Inorganic scintillators contains large fraction of elements with high Z (e.g. an effective atomic number of LYSO crystal is equal to 66 u [14]) and attenuation length in that type of scintillators is short.

The vast majority of inorganic scintillators are crystals. The mechanism of scintillation is based on electron-hole pairs production in the valence and conduction band during interaction with incident radiation. Light output of inorganic scintillators can be higher (see Table 2) in comparison to organic ones. However they are expensive and the process of crystal growth is difficult to carry out [33].

Organic scintillators are built of chemical substances including phenyl rings. They are found in three types: crystalline, like anthracene or stilbene, liquid, when the scintillator is dissolved in solvent e.g. xylene or toluene and plastic scintillators. Crystalline organic scintillators are expensive and vulnerable. Liquid ones are toxic and its utilization is inconvenient. Because of the volatility, they need to be stored in special containers.

The mechanism of luminescence in organic and inorganic crystals differs significantly, what is determined by their intrinsic structure. In organic crystals, molecules are weakly bounded in comparison to inorganic compounds. In such loose arrangement energetic levels are not disturbed by the environment [34].

This thesis concerns plastic scintillators. The scintillation mechanism of organic scintillators will be exemplified for plastic scintillators in the following chapter.

## 3.1. Scintillation process

The mechanism of scintillation in plastic scintillators is fluorescence. In Fig. 4 the typical energy level diagram in organic scintillators is shown.

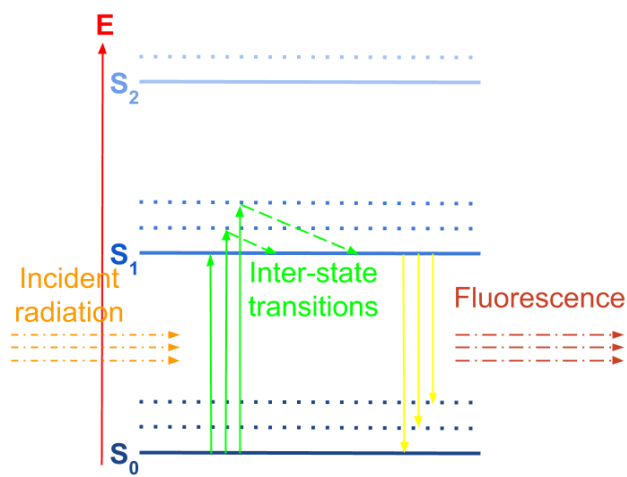

**Figure 4 Scheme of the singlet energy level diagram of organic scintillators [33].**

Ionising radiation, incident on the scintillator is partially absorbed and emitted via fluorescence. In this process, the absorbed energy is instantaneously converted to the emitted energy - the decay time in organic scintillators is equal to $10^{-10}$ - $10^{-7}$ s. Fluorescence bands are placed within larger wavelengths in comparison to the incident radiation because emission transition occurs after releasing a part of oscillation energy to the surroundings.

Molecules have particular system of energy levels (see Fig. 4): electron, oscillation and rotation. Absorbing the energy, the molecule proceeds to one of the excited states. The excess of the energy can be lost in few ways.

The energy of incident radiation is transferred to particular atoms, causing an electron transition from the basic $S_0$ state to the excited state $S_1$ or higher, depending on the energy. Non-radiative transitions occur quickly between vibrational states of $S_1$ (green dashed lines in Fig. 4). Electrons fall from $S_1$ vibrational states to $S_1$ basic state is favourable for scintillators. Then, there is an electron transition from $S_1$ to $S_0$ state. The excess of the energy is radiated as fluorescence photons within UV or visible wavelengths.

For scintillation detectors, the transition from $S_1$ vibrational states to $S_1$ base level is favourable. In such decay electrons loose a part of energy. This is a reason of difference between the absorbed and emitted energy. As a consequence, the absorption and emission spectra of scintillating materials are shifted as a function of light wavelength. The difference between the wavelength of the maximum of absorption and the maximum of fluorescence is called Stokes shift. Scintillators are materials with particularly large Stokes shift and therefore the re-absorption of scinitllating light is unlikely [33].

The base of plastic scintillator is polymerizable liquid like styrene or vinyltoluene. This substances scintillate in UV, however the mean free path of produced photons is too short. Therefore scintillation additives are used. Additives (fluors) absorb the primary light from the base and emit it in longer wavelengths. One or more fluors can be used, depending on the desired wavelength of emitted photons.

Typical plastic scintillators are ternary systems, consisting of three components: polymeric base, primary fluor and secondary fluor, so called wavelength shifter (WLS) [35]. The scheme of mechanism of the energy transfer in plastic scintillator is shown in Fig. 5. Three components of the plastic scintillators are shown in block scheme in particular sequence, however the scintillator is a homogeneous mixture of these chemical compounds.

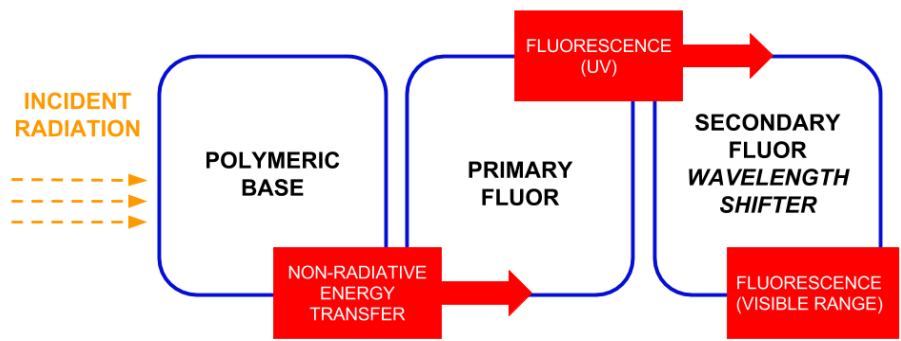

**Figure 5 Block scheme of energy transfer in plastic scintillator.**

The incident radiation interacts with polymer molecules exciting them. The energy is transferred in non-radiative way to primary fluor through Förster mechanism. This is a process occuring in excited states, when the emission spectrum of the donor fluorophore overlaps with the absorption spectrum of the acceptor - primary fluor. In case of polystyrene and polyvinyltoluene, fluorophores are the delocalized π electrons. The light emission of the donor is not involved. Förster energy transfer is based on dipole - dipole interaction between molecules of donor and acceptor distant from each other by 30 - 60 Å [36].

Primary fluor absorbs the energy and emits it in UV range via fluorescence (see Fig. 5). This wavelength is not adjusted to the quantum efficiency of the light detectors which are photomultipliers. To shift the maximum wavelength of emission towards larger wavelengths, wavelength shifter is used. This substance absorbs the light emitted by primary fluor, and as a result, photons in visible range are produced. Such light can be efficiently detected by a photomultiplier.

## 3.2. Chemical compounds of plastic scintillators

Plastic scintillators are obtained by polymerization of the liquid monomer in which scintillating additives are dissolved. In effect, a block of homogeneous scintillator can be obtained. Nowadays, polyvinyltoluene (PVT) is the most widely used matrix for plastic scintillators base because of the best scintillating properties among polymers. Commercially available scintillators offered by Eljen Technology [13] or Saint-Gobain [12] are based on polyvinyltoluene. Polystyrene (PS), its homolog, is very popular as scintillator matrix as well [37]. Chemical structures of this compounds are shown in Tab. 3.

Table 3 Chemical structures of polymeric matrix of plastic scintillator [33].

| Name | Chemical structure |
|---|---|
| Polystyrene (PS) | 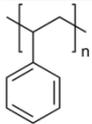 |
| Polyvinyltoluene (PVT) | 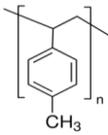 |

Both monomers: styrene and vinyltoluene are viscous liquids which polymerize on heating to form of solid, thermoplastic polymers. The basis of all aromatic vinyl polymers, including polystyrene and polyvinyltoluene, are -CH=CH$_2$ bounds which breaking enable to built long polymeric chains [37].

Polymer cannot be an effective scintillator because of weak fluorescence efficiency and short mean free path for scintillating light. In molecules containing aromatic rings, only small fraction, about 3 % of the energy is converted into optical photons [38]. The wavelength of emitted photons is not adjusted to quantum efficiency of photomultiplier (see Fig. 5). Polymer is just the medium to transfer the energy to primary fluor. About 1 wt. % of the dopant increases not only the attenuation length of photons but also the scintillator light output [33].

There is a large number of chemical compounds, that can be used as primary fluors in plastic scintillators. Some of them are presented in Tab. 4.

Table 4 Substances that can be used as primary fluors in plastic scintillators. $\lambda_{abs}$ denotes wavelength at maximum of absorption, $\lambda_{em}$ - wavelength at maximum of emission, $\Phi_f$ - fluorescence quantum efficiency, $\tau$ - decay time of light pulses, R - solubility in a given solvent at 20°. Denotation of solvents: mb– toluene, a – acetonitryle, b – benzene, c – cyklohexsan, d – dioxane, e – ethanol, k – xylene, m – methylene chloride, me – methanol, t – THF. Data is taken from [36] [39] [40] [41] [42] [43].

| Chemical structure, name | Abbreviation | $\lambda_{abs}$ [nm] | $\lambda_{em}$ [nm] | $\Phi_f$ | $\tau$ [ns] | R [g/dm$^3$] |
|---|---|---|---|---|---|---|
| 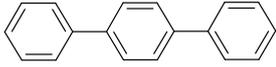 p-terphenyl | PTP | 288mb 276c | 335mb 339c | 0.85mb | 1.2mb 1.05e 1.16me 0.99c | 8.6mb 6k |
| 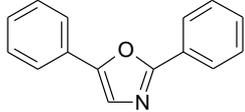 2,5-diphenyloxazole | PPO | 308mb 303e | 365mb 361e 375d | 0.8mb | 1.6mb 1.3b 1.35c 1.4e | 414mb 335k |
| 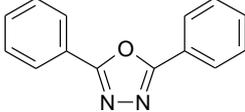 2,5-diphenyl-1,3,4-oxadiazole | PPD | 283mb | 355mb | 0.9mb | 1.5mb 1.2e | 70mb |
| 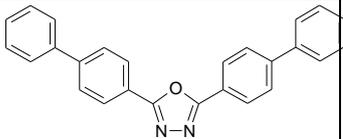 2,5-bis(4-biphenyl)-1,3,4-oxadiazol | BBD | 314d 315mb | 373d 380mb | 0.85mb | 1.4mb | 2.5mb 1k |
| 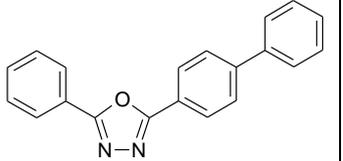 2-phenyl-5(4-biphenyl)-1,3,4-oxadiazole | PBD | 305mb 302e | 360-356mb 362e | 0.8mb | 1.2mb | 21mb 18k |
| 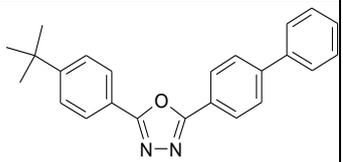 2-(4-tert-buthylphenyl)-5-(4-biphenylo)-1,3,4-oxadiazole | BPBD | 308mb | 365mb 368c | 0.85mb | 1.2mb | 119mb 77k |

Decay times of light pulses in all substances from Tab. 4 are similar. However they have different maxima of emission wavelength, so one can adjust the substance to particular application. Solubility in a given solvent is very important parameter because only complete dissolution enables proper scintillator acting.

There are also large number of substances that can be used as wavelength shifters in plastic scintillators. Some of them are presented in Tab. 5.

**Table 5** Substances used as wavelength shifters in plastic scintillators. The source of data and used notations are the same as in Table 4.

| Chemical structure, name | Abbreviation | $\lambda_{abs}$ [nm] | $\lambda_{em}$ [nm] | $\Phi_f$ | $\tau$ [ns] | R [g/dm³] |
|---|---|---|---|---|---|---|
| 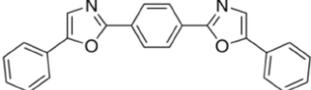 1,4-bis(5-phenyl-2-oxazolyl)benzen | POPOP | 365mb 358c 358e | 415-417mb 410c 425e | 0.85mb 0.97c | 1.13c 1.5mb 1.35e | 2.2mb 1.4k |
| 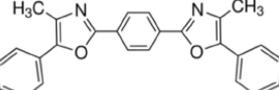 1,4-bis(4-methyl-5-phenyl-2-oxazolyl)benzene | DM-POPOP | 370mb | 430mb | 0.93mb | 1.5mb 1.45e | 3.9mb 2.6k |
| 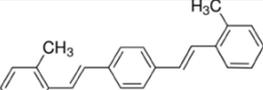 1,4-bis(2-methylostyryl)benzene | Bis-MSB | 347mb 350c | 418e 420mb 420c 423d | 0.96c | 1.6c | No data |
| 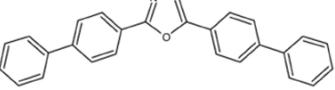 2,5-di(4-biphenylo)oxazole | BBO | 340mb 342b | 410-412mb 409b | 0.75mb | 1.4mb | 3.1mb 1.3k |
| 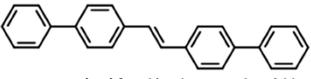 trans-4,4'-diphenylstilbene | DPS | 337mb 340d 341b | 410mb 408b | 0.8mb 1.0mb | 1.1mb | 1.5mb 0.9k |
| 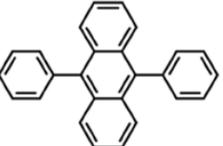 9,10-diphenylanthracene | DPA | 366-375c | 430c | 0.95-1.0c | 8.7me 7.3c 9.35c | 35mb |

One can notice that solubility of wavelength shifters in given solvents is rather weak. However, the amount of WLS in plastic scintillator is about 1 ‰, thus dissolution of such portion is possible.

## 3.3. Commercial plastic scintillators

There are many worldwide companies producing plastic scintillators e.g. Saint Gobain [12], Eljen Technology [13] or Rexon [44]. Headquarters and production lines of these companies are placed in United States. In Europe, there are two companies developing plastic scintillators: Nuvia in Czech Republic [45] and Amcrys in Ukraine [46]. Moreover, there is an Epic Crystal company in China [47] which also produces plastic scintillators.

Plastic scintillators offered by the European and Chinese companies are based on polystyrene. This polymer does not work as efficiently as polyvinyltoluene as a scintillator matrix, decreasing the scintillators performance by 15 - 20 % [48]. Because of that, the best scintillators should be purchased in USA, however this is connected with additional delivery charges and duty.

Development in the field of photoelectric converters, e.g. replacing of vacuum photomultipliers by silicon ones, forces improvements of plastic scintillator to match them to the new photomultipliers, in order to construct efficient detectors. This includes the increase of wavelength of scintillation light, which is justified by better propagation of light in the scintillating material and matching to the SiPM quantum efficiency wavelength dependence. Development of fast signals processing electronics opens possibility of using scintillators with short rise time of signals.

Worldwide manufacturers offer plastic scintillators with different values of light output, light attenuation length, rise and decay times of signals, and different emission spectra. Planning an experiment, one can chose the best scintillator for a particular application.

However, vast majority of plastic scintillators offered by companies is well adjusted to traditional vacuum photomultipliers, popular by many years. The demand of SiPMs is to shift the emission spectrum of scintillators towards longer wavelengths.

Market of plastic scintillators is widely opened for the new solution. For example, slogan of Rexon is: "IF WE DON'T HAVE IT, WE'LL MAKE IT". The company offers designing and manufacturing of plastic scintillators to the particular need, concerning scintillator's composition and shape. However, it is much better to have developed and tested scintillator because the elaboration of the novel scintillator is a time-consuming and labour - intensive process. New scintillators are utilized usually in the laboratory experiments, having financial support given for the particular time of the project realization. That is why, it is important to have the new scintillator available and fully characterized, to enable its immediate use for building detectors.

Because of specific needs of plastic scintillators market, in this thesis a novel solution was proposed. Novel plastic scintillator was elaborated and tested. Its novelty lies in the use of 2-(4-styrylphenyl)benzoxazole as a wavelength shifter. The scintillator can be used in many experiments of physics.

## 3.4. Applications of 2-(4-styrylphenyl)benzoxazole

In this thesis for the first time application of 2-(4-styrylphenyl)benzoxazole as a plastic scintillator dopant is described. The aim of the dopant is shifting of the scintillator's emission spectra towards longer wavelengths to match it to the quantum efficiency of SiPMs.

2-(4-styrylphenyl)benzoxazole can be obtained in relatively simple and cost-effective chemical synthesis. Method of synthesis is described in patent [49] and in the review article [50]. The substance was synthesized for the first time by A. E. Siegrist and collaborators in 1960 and applied as optical brightener.

2-(4-styrylphenyl)benzoxazole was used also as an emission material in OLED (organic light - emitting diode), electroluminescent diodes [51] or as NLO (nonlinear optics) material [52]. It was applied in cosmetology, dentistry and in photographic materials [53]. The substance was utilized in photosensitive coating [54] as well as a whitening agent [55] and in optical disks for lasers [56].

## 4. Experimental methods

Polymerization process, in which plastic scintillators are obtained is conducted in a furnace in a specially designed glass reactors or in a form. Scintillating dopants were dissolved in liquid purified monomer (styrene or vinyltoluene). Scintillators were obtained by bulk polymerization of such prepared samples. The mechanism of this kind of polymerization is free radical. This is a process occurring in pure monomer. To avoid contamination of the material, the polymerization is thermally initiated, without any chemical initiators. High concentration of monomer enables high rates and degrees of polymerization. However, there is a problem with increasing viscosity of the mixture when the polymerization proceeds. Bulk polymerization reaction is highly exothermic and increasing viscosity inhibits heat flow leading to formation of regions of local overheating. As a result, empty voids can be generated in a block of polymer because of the internal shrinkage.

Bulk polymerization allows to produce scintillator characterized by high light output due to e.g. its homogeneity [57]. The temperature schedule was adjusted to obtain optically homogeneous scintillator samples and eliminate effects of polymerization shrinkage. Production of scintillators and optimization of their composition maximizing scintillator light output was the first stage of the research.

Two scintillating dopants were dissolved in the monomer: primary and secondary fluor. In prepared scintillators, primary fluor is a commercially available compound: 2,5-diphenyloxazole (PPO). As a secondary fluor 2-(4-styrylphenyl)benzoxazole was used. It is a substance chosen amongst three chemical compounds which was synthesized and tested as wavelength shifter in plastic scintillator. Only 2-(4-styrylphenyl)benzoxazole posseses exceptionally good scintillating properties. The use of this substance as a scintillator dopant and the novel scintillator composition are subject of patent application [58].

Measurements of light yield were carried out in detector laboratory. Charge spectra were registered irradiating scintillators with $^{22}$Na source of gamma quanta with energy of 511 keV originating from annihilation of positron with electron (Fig. 1). The source was placed in the lead collimator providing narrow beam of gamma quanta, about 1 mm wide.

Interaction of gamma quanta with the scintillator results in production of light. To both sides of scintillator photomultipliers are connected (see Fig. 2). They play a role of converters of scintillation light into electrical signals. Then signals are collected and processed by the oscilloscope. Determining the position of the middle of the Compton edge on the charge spectrum histogram, light output of manufactured and purchased scintillators were appointed and compared. Light output is the most important parameter of the scintillator and determines the number of photons emitted per unit of energy deposited in the scintillator. It gives information about the effectiveness of conversion of the incident radiation into photons.

Optimal concentration of the novel scintillating dopant: 2-(4-styrylphenyl)benzoxazole was set maximizing light output of the scintillator. Scintillators with different concentrations of the dopant were prepared and light output of the samples were measured.

Based on measurements conducted with the setup enabling determination of the light output, characterization of signals arising in synthesized and commercial scintillators were done. Rise and decay times of signals were determined and compared with commercially available scintillators. The shorter the decay time, the better is the scintillator concerning application in J-PET/MR scanner.

Scintillators were subject of tests in order to measure the emission spectrum. Emission spectra of thin samples of J-PET were registered and compared with quantum efficiency of silicon photomultipliers to check if they are matched to each other. A proper matching of these quantities is necessary for an effective light conversion into electrical pulses by photomultiplier.

Characterization of scintillators structure by analyzing sizes and fraction of free volumes in particular samples were carried out using Positron Annihilation Lifetime Spectroscopy (PALS). This technique enables very accurate analysis of free volume sizes in the scintillator and any structural transitions occurring with the temperature changes. Glass transition temperature and temperatures of some structure changes correlated to organization of molecules can be observed.

Samples of plastic scintillators were analyzed also using Differential Scanning Calorimetry (DSC). This is the method which can be considered as a complementary to PALS. It is based on measurement of the amount of heat released or absorbed during physical or chemical process. DSC allows to describe thermal transitions in polymers. In case of polystyrene or polyvinyltoluene, the most significant is glass transition ($T_g$) temperature. It is a point of transition of amorphous brittle polymer into rubbery one. $T_g$ is also visible in PALS measurement, therefore results obtained in both methods were compared.

# 5. Potential wavelength shifters for the J-PET scintillator

Scintillators offered by companies are mostly adjusted to vacuum photomultpliers which were widely used in physics experiments for last decades. However, progress in the area of photodetectors, and successive substitution of traditional photomultipliers by SiPMs, demands development in the area of scintillators as well. Development of novel scintillator with the emission spectrum adjusted to the novel J-PET/MR device, enables an efficient detection of gamma quanta.

In order to develop such scintillator, three chemical compounds were synthesized. The syntheses are described below in subsections A, B and C.

A) Synthesis of 2-(4′-*N*,*N*-dimethyloaminophenyl)oxazolo[4,5-*b*]pirydyne (DMAPOP)

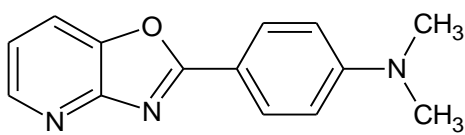

DMAPOP was synthesised according to procedure described in [59]. The scheme of reaction is presented below:

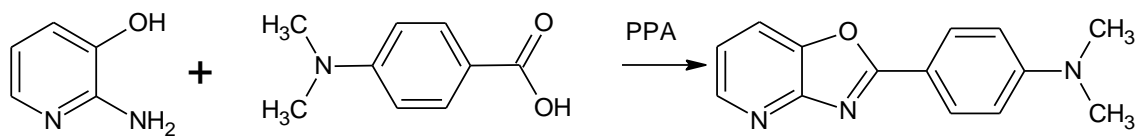

PPA - polyphosphoric acid

In order to purify the desired compound, it was recrystallized from chloroform and then the final product was separated by column chromatography with Al$_2$O$_3$ as stationary phase and chloroform as eluent. Then the sediment was precipitated from petroleum ether.

The next step was column chromatography on silica gel/CHCl$_3$. Fractions containing desired compound were selected according to results of TLC (Thin Layer Chromatography). The chosen fractions were connected and the solvent was evaporated. The sediment was precipitated to petroleum ether. The chemical compound with high purity was obtained, which melting temperature was equal to 227 °C.

B) Synthesis of 2-(4-styrylphenyl)benzoxazole

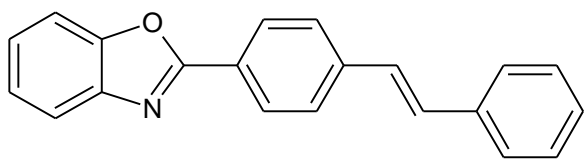

The reaction proceeds according to the scheme:

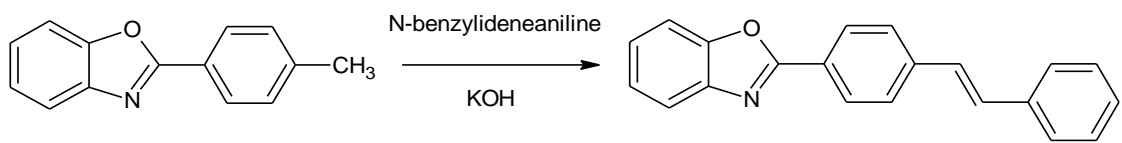

The final product was synthesized. Obtained solid was dissolved in toluene and particular fractions were separated using column with silica gel. Then with TLC technique proper fractions were chosen, and the sediment was precipitated to petroleum ether. Melting temperature of the substance was 198 °C.

C) Synthesis of 2-[4-methylbenzene][1,3]-oxazolo[5,4-*b*]-quinoline.

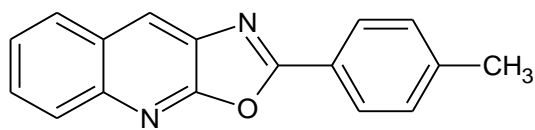

The reaction was as follows:

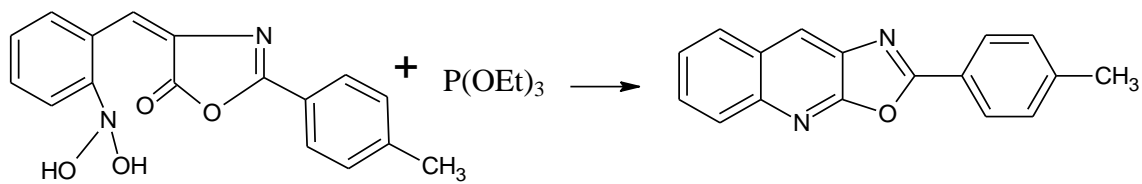

P(OEt)$_3$ - triethylphosphine

The product of the reaction was dissolved in toluene and particular fractions were separated in column filled with Al$_2$O$_3$. Because of the insufficient degree or purity confirmed by TLC, the sediment was dissolved in chloroform and components of the mixture were separated by column with silica gel and toluene as eluent. Proper fractions were selected by TLC and solvent was removed by rotor evaportor. The sediment was crystalized from toluene. Melting point of the reaction product was equal to 243 °C.

In scintillators development, high degree of purity is demanded. Because of that, each synthesis was followed by purification process. High purity of reaction products was confirmed by measurements of the melting points.

Plastic scintillators containing synthesized fluorophores as wavelength shifters were prepared. Their constitution is shown in Tab. 6.

Table 6 Constitution of plastic scintillators containing new wavelength shifters.

| Polymeric base | Primary additive | Secondary fluor |
|---|---|---|
| Polystyrene | p-terphenyl (PTP) / 2,5-diphenyloxazole (PPO) 1 % | Synthesized substances in amount 0.01 % |

Scintillators were cut and polished to a geometry of cylinder with diameter and height equal to 2.5 cm. They were wrapped with teflon foil and their interaction with gamma quanta from $^{68}$Ge source was tested. A simple experimental setup consisting of vacuum photomultiplier, scintillator and source was built (Fig. 6) and charge spectra were registered.

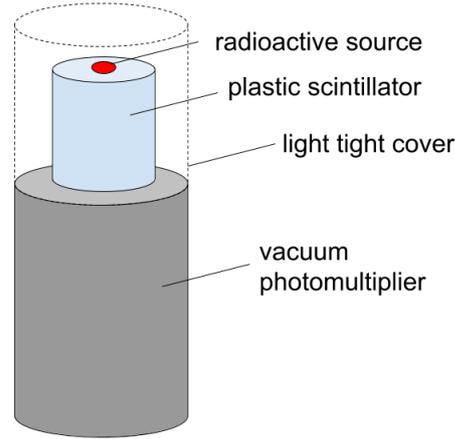

**Figure 6 Experimental setup for the estimation of plastic scintillators light output.**

Detailed analysis of spectra registered during gamma quanta interaction with plastic scintillator and description of the phenomenon itself are the subject of Chapter 7.2: Spectral properties of J-PET scintillators. The first preliminary studies relied on fitting Fermi function to the falling edge of the spectrum. That allows to estimate roughly the relative light yield of tested scintillators. Fermi function is given by the equation (1):

$$f_F(x) = \frac{P_0}{e^{\frac{x-P_1}{P_2}} - 1} + P_3 \quad (1).$$

Parameters denoted as $P_0$, $P_1$, $P_2$, $P_3$ are related to maximal values of the function, the middle of the edge, slope and minimal values of the function, respectively.

The function was fitted to spectra as it is shown in Fig. 7 registered for stilbene which was taken as standard. In the spectrum, histogram of signals amplitudes is presented. Amplitudes are proportional to the number of photons produced by scintillator during interaction with gamma quanta. The absolute number of photons produced by scintillator interacting with gamma quanta was not calculated. The aim of this test was to check if developed materials have a potential for being used as scintillators.

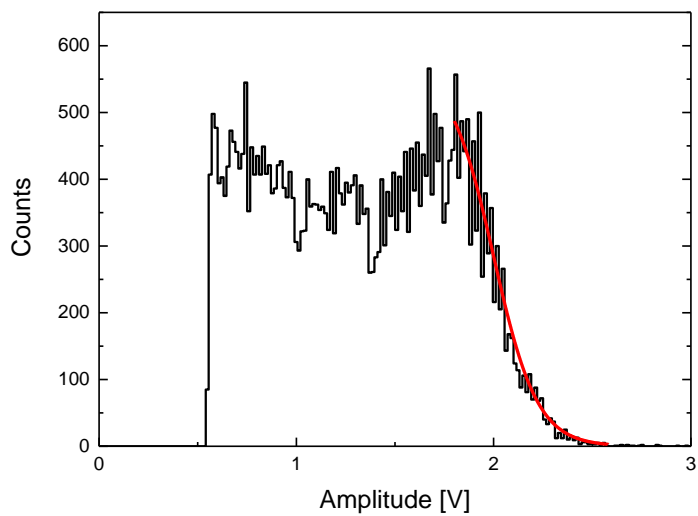

**Figure 7 Amplitude spectra of stilbene with Fermi function fitted (red).**

$P_1$ parameter describing the middle of the falling edge, was a measure of scintillators light output. $P_1$ determined for obtained scintillators were compared to the parameter value of stilbene. Results of such rough estimation of scintillators light output are presented in Tab. 7.

**Table 7 Relative light output of tested plastic scintillators with respect to the light output of stilbene.**

| Scintillator | Relative light output |
|---|---|
| Stilbene | 100 % |
| Plastic scintillator containing 2-(4′-*N*,*N*-dimethyloaminophenyl)oxazolo [4,5-*b*]pirydyne | ~ 20 % |
| Plastic scintillator containing 2-(4-styrylphenyl)benzoxazole | ~ 80 % |
| Plastic scintillator containing 2-[4-methylbenzene][1,3]-oxazolo[5,4-*b*]-quinoline | ~ 50 % |

Light output of plastic scintillators obtained in laboratory conditions is equal to about 20 %, 80 % and 50 % of stilbene, respectively. Because of the highest value of light output amongst all tested samples, scintillators containing 2-(4-styrylphenyl)benzoxazole are a subject of further tests described in this thesis.

## 6. Development of plastic scintillators

The most popular method of plastic scintillators development is bulk polymerization. This is the method widely used for polymerization of all vinyl monomers. Liquid monomer: styrene, vinyltoluene or methyl methacrylate with proper additives dissolved therein can be polymerized in this way [60]. Bulk polymerization is a homogeneous process which is conducted in the ambient of pure monomer without any additional solvents. It is possible to use substances that initiate the polymerization reaction which are called initiators. However, in synthesis of plastic scintillators, the high degree of purity is required, so it is better to avoid any additional substances. Moreover, styrene and its homologs can be their own initiators when the temperature is raised. The reaction is thermally initiated, therefore the process is called thermal polymerization.

The mechanism of that reaction is free radical. Styrene is one of well known monomers prone to spontaneous generation of free radicals by itself at elevated temperatures without any chemical initiators. Two most famous: Mayo and Flory mechanisms are presented in Fig. 8.

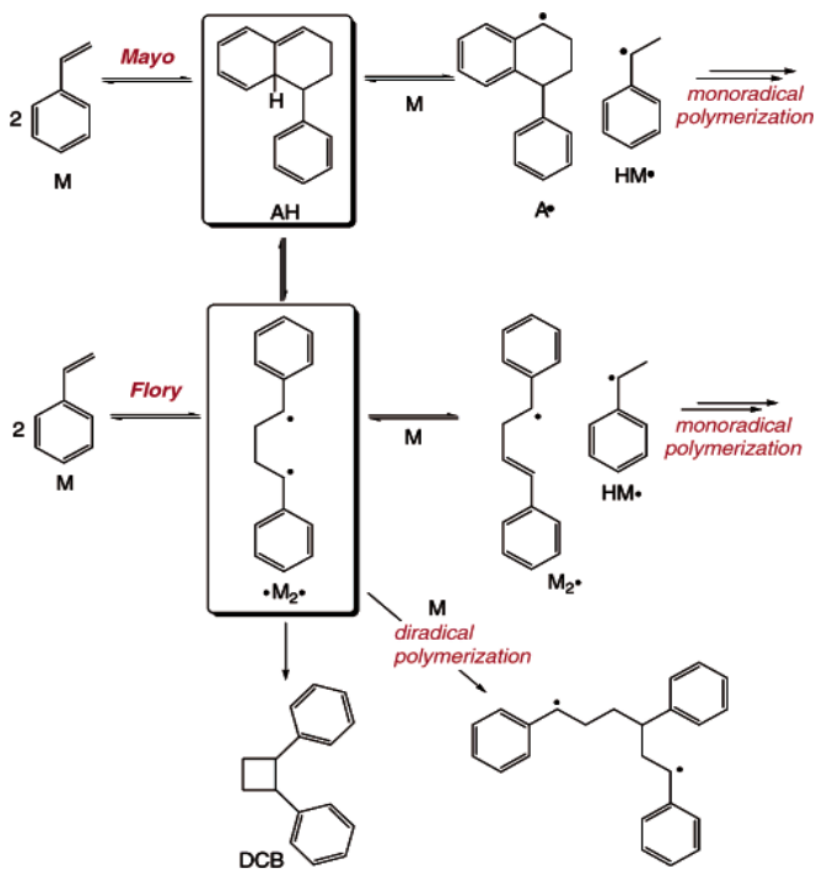

**Figure 8 Mayo's and Flory's mechanisms of free radical generation in styrene polymerization. The scheme is taken from the article [61].**

In the Mayo mechanism, the polymerization is initiated by Diels-Alder dimerization of styrene. Because of homolysis between the dimer (AH) and a third styrene, monoradicals A• and HM• are generated. HM• initiates the polymerization.

In Flory's interpretation, styrene dimerizes to 1,4-diradical (M2). A third styrene abstracts a hydrogen atom from the diradical and monoradical initiators are formed.

In both: Mayo and Flory mechanisms the diradicals may form 1,2-diphenylocyclobutane (DCB) which is not capable of initiating polymerization, or AH dimer, what leads to Diels-Alder cycloadduct production and in turn initiates the polymerization reaction [61].

The further steps of polymerization following initiation, are propagation and termination. Propagation of polymeric chain can be schematically described as:

$R_n\bullet + M \rightarrow R_{n+1}\bullet$

where $R_n\bullet$ denotes polymer chain with n mers, M – monomer, $R_{n+1}\bullet$ – polymer chain extended by one mer.

Termination can occur by combination:

$R_x\bullet + R_y\bullet \rightarrow R_{x+y}$

or disproportionation:

$R_x\bullet + R_y\bullet \rightarrow R_x + R_y$ [62].

Free radicals are highly reactive. A radical attacks a molecule of monomer, initiating a macroradical, to which other monomer units are added successively. Each time the active centre is transferred to the attached mer. This is the way in which propagation of polymer chain occurs.

As the reaction proceeds, the amount of monomer decreases, interpolymeric reactions start to occur. Active parts of macroradicals meet each other and termination takes place. It leads to dead polymer and may occur via combination or disproportionation.

The main disadvantage of this kind of polymerization are problems with the dissipation of the exothermicity in the sample volume. In effect, local hot spots are formed [63]. During free radical polymerization, an autoacceleration occurs, which is known as Trommsdorff or gel effect [64]. This is a phenomenon of spectacular increase of polymerization rate. When the reaction speeds up, viscosity grows and huge amounts of heat that are released accelerate this reaction further [65].

Trommsdorff effect occurs at high conversion, when termination reactions dominate and "dead" polymers, without active parts are produced. The vast majority of polymer chains are long and slightly mobile. They are so immobilized that their fastest termination is with short, mobile, unentangled chains. Gel effect leads to uncontrolled local rise of temperature what may result in the reactor explosion [66].

The other disadvantage of the bulk polymerization is that during the process the polymerization shrinkage occurs. The most of the shrinkage appears in the early stages of polymerization. It covers about 20 per cent of the initial volume. The shrinkage is a cause of cracks and many other flaws like empty void in a block of polymer [67]. This is very disfavourable in case of plastic scintillators because optical homogeneity of the material is essential to provide the proper performance of the detectors.

However, conducting scintillators synthesis with the temperature schedule which is presented in Fig. 9 enables to obtain samples without defects. Firstly the reaction mixture is quickly heated to 100 °C and then to 140 °C within few hours. Next, the principal heating occurs and lasts about 70 hours. After this time, monomer is reacted into polymer and the sample is cooled slowly to 90 °C. The scintillator is stored in this temperature for three hours in order to anneal the sample. Then the scintillator is finally quenched.

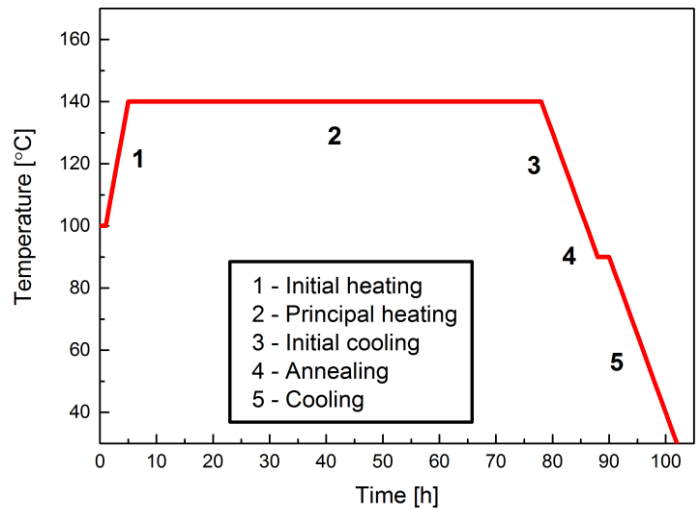

**Figure 9 Temperature schedule of the plastic scintillators polymerization [68].**

Scintillators are prepared by dissolving scintillating additives in liquid monomer: styrene or vinyltoluene. As the primary fluor 2,5-diphenyloxazole (PPO) or p-terphenyl (PTP) were used in an amount of 2 wt. %. As a wavelength shifter 2-(4-styrylphenyl)benzoxazole was applied. The concentration of the novel wavelength shifter was optimized as it is described in Chapter 10: Light output of J-PET scintillators.

The reaction mixture was poured into glassy ampoules, inert gas purged and then tightly sealed in the flame of the burner. Such prepared samples had been placed in the tube furnace and then polymerization was started according to the schedule shown in Fig. 9. To minimize the risk of explosion, the mixture occupies the half of the ampoule volume at maximum. Photographs of obtained scintillators are shown in Fig. 10.

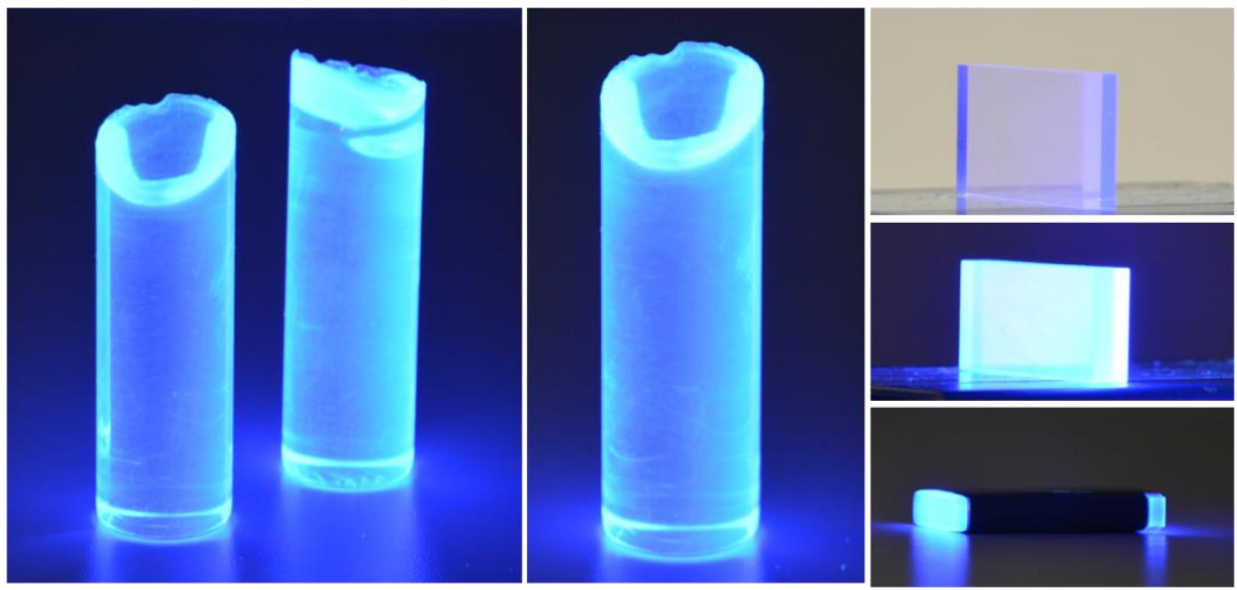

**Figure 10** J-PET plastic scintillators in UV light. Left and centre: scintillators removed from the glass ampoule. Right from the top: scintillator cut and polished, exposed to UV in daylight, in the dark and scintillator wrapped in light tight foil in the dark.

## 7. Optical, spectral and timing properties of J-PET scintillator

### 7.1. Optical properties of J-PET scintillator

In order to characterize properties of the scintillator containing the novel wavelength shifter, and check if it is matched to spectral properties of silicon photomultipliers, absorption and emission spectra of the wavelength shifter: 2-(4-styrylphenyl)benzoxazole and emission spectra of the scintillator were registered.

Absorption and emission spectra of $4.2*10^{-8}$ M 2-(4-styrylphenyl)benzoxazole solution in toluene were registered. They are shown in Fig. 11. Maximum of absorption lies at 349 nm, while the emission spectrum is characterized by the maximum in 402 nm.

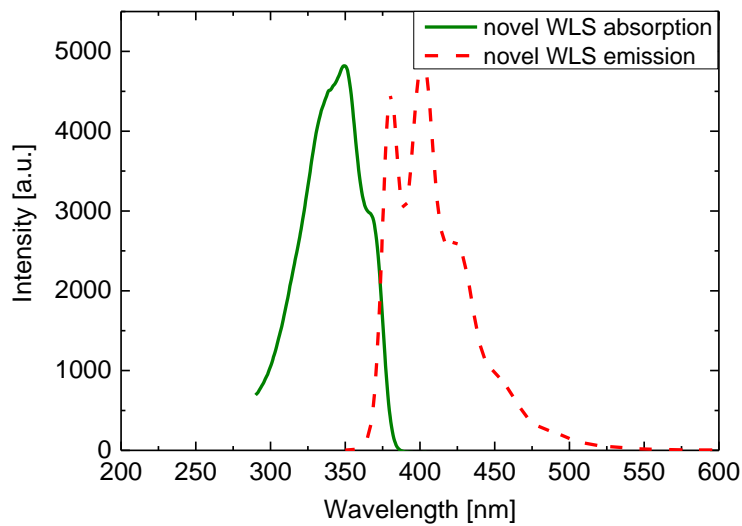

**Figure 11 Absorption and emission spectra of 2-(4-styrylphenyl)benzoxazole. Absorption spectrum (solid green line) was registered with Hitachi U-2900 Spectrophotometer, and emission spectrum (red dashed line) by Hitachi Fluorescence Spectrophotometer F 7000, the plot was corrected by the detector sensitivity.**

The energy absorbed and emitted by particular dyes is associated with electron transitions to different vibrational states. According to Fig. 4, the energy of excited electrons is lost in the transitions from $S_1$ vibrational states to $S_1$ basic state and from $S_1$ ground level to $S_0$. The energy emitted in these processes is less than the energy which was absorbed during electron excitation from $S_0$ to $S_1$. As a result, spectra of absorption and emission are shifted relatively to each other what reduces re-absorption. This phenomenon is based on Stokes law, which implies that the wavelength of emission was greater than the wavelength of absorption. The difference between maxima of absorption and emission peak is identified as Stokes shift. The larger the Stokes shift is, the smaller probability of re-absorption of scintillating light in the material. Re-absorption is undesired because it shortens the attenuation length. That is why substances with large Stokes shift are valuable scintillator dopants [33] [38].

In Fig. 12 absorption and emission spectra of POPOP (1,4-bis(5-phenyloxazol-2-yl)benzene) which is widely used wavelength shifter in plastic scintillators are shown. The value of the Stokes shift which is calculated as the difference between maxima of absorption and emission, is equal to about 50 nm.

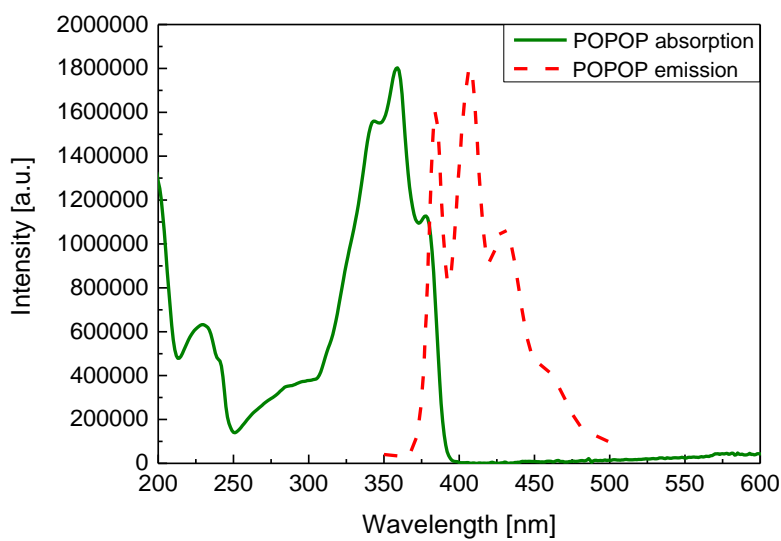

**Figure 12 Absorption (green solid line) and emission (red dashed line) spectra of POPOP in cyclohexane [69].**

In the case of 2-(4-styrylphenyl)benzoxazole, the Stokes shift resulting from the difference between peaks of 402 nm and 349 nm, is equal to 53 nm. This value is comparable to Stokes shift of POPOP. Therefore Stokes shift of 2-(4-styrylphenyl)benzoxazole is sufficient to use the substance as a scintillator dopant.

Another issue which is necessary to consider is adjustment of the primary additive to the 2-(4-styrylphenyl)benzoxazole to provide an effective energy transfer. According to Fig. 5, in plastic scintillator the energy is transferred from primary to secondary fluor in the way of fluorescence. However to provide an effective transfer of energy, emission spectrum of primary fluor has to overlap with absorption spectrum of secondary fluor. In Fig. 13 and Fig. 14 absorption and emission spectra of plastic scintillating additives are presented. As a primary additive 2,5-diphenyloxazole (PPO) and p-terphenyl (PTP) were considered while 2-(4-styrylphenyl)benzoxazole (novel WLS) is assumed to act as a wavelength shifter, which is secondary additive.

In both sets of mixture: PPO + novel WLS and PTP + novel WLS, the emission spectrum of the primary fluor overlaps with the absorption spectrum of secondary fluor. In case of PPO, their maxima fits almost perfectly to each other, while PTP and novel WLS maxima are slightly shifted within few nanometers.

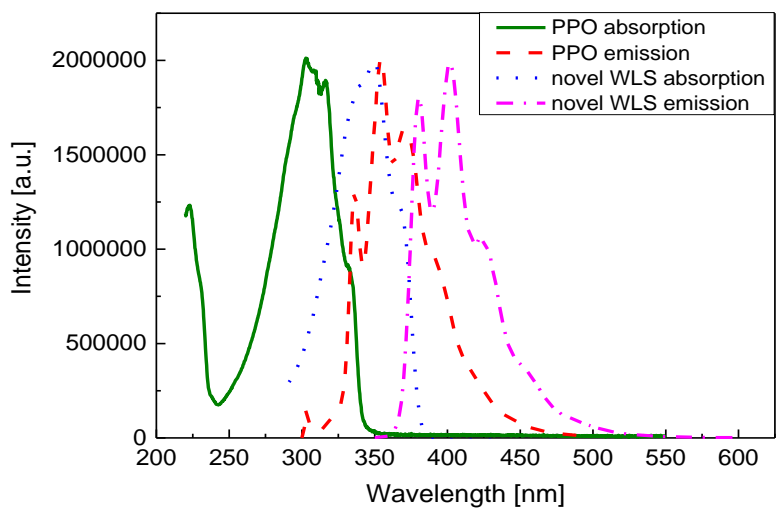

**Figure 13** Absorption and emission spectra of 2,5-diphenyloxazole (PPO) are marked with green solid line and green dashed line. Spectra are taken from [69]. Absorption and emission spectra of 2-(4-styrylphenyl)benzoxazole (novel WLS) are marked with blue dotted line and pink dashed-dotted line, respectively.

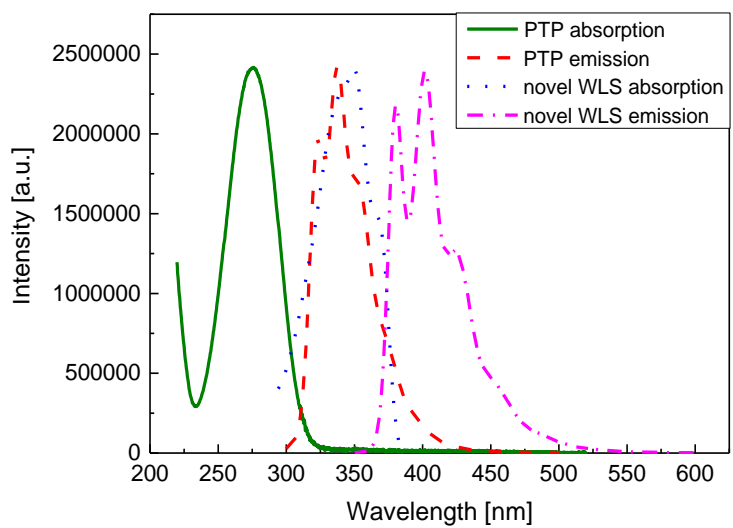

**Figure 14** Absorption and emission spectra of p-terphenyl (PTP) are marked with green solid line and red dashed line, respectively. Spectra are taken from [69]. Absorption and emission spectra of 2-(4-styrylphenyl)benzoxazole (novel WLS) are marked with blue dotted line and pink dashed-dotted line, respectively.

Emission spectra of J-PET and BC-420 scintillator were registered with FluoroLog-3 (Horiba Jobin-Yvon) in reflecting mode using PMT R928P and continuous wave xenon source. The obtained emission spectrum of the J-PET scintillator is indicated as green solid line in Fig. 14. It is clearly visible that J-PET spectrum is broader than BC-420 (red dashed line). This means that photons are emitted by the J-PET scintillator in larger wavelength range than photons emitted by BC-420.

In Fig. 15 quantum efficiency of vacuum and silicon photomultipliers are presented as well. Broadening of J-PET scintillator spectrum towards longer wavelengths enables more effective registration of its scintillation light by SiPMs. Emission spectrum of J-PET scintillator overlaps the region of larger SiPMs quantum efficiency in comparison to BC-420. Quantum efficiency of SiPMs is larger than quantum efficiency of vacuum photomultipliers and it increases significantly with the increasing wavelength (up to about 450 nm) as indicated by pink dashed-dotted line in Fig. 15.

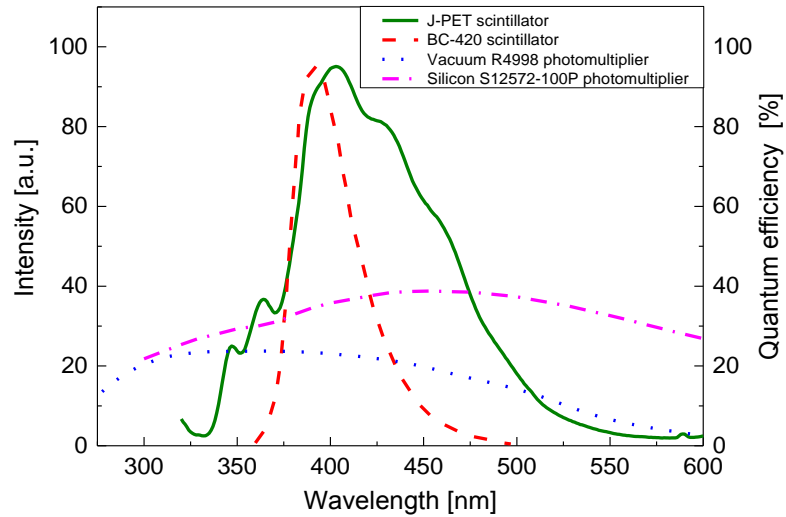

**Figure 15 Emission spectra of J-PET (green solid line) and BC-420 [12] (red dashed line) scintillators superimposed on the quantum efficiency dependence on photons wavelength for typical vacuum tube photomultiplier with bialcali window (blue dotted line) [21] and silicon photomultipliers (pink dashed-dotted line) [21]. Maximum of emission for BC-420 scintillator is placed at wavelength of 393 nm, while the maximum of J-PET scintillator at 403 nm. The emission spectra are normalized in amplitude.**

In the article [70] simulation of photon transport in cuboidal scintillators was described. Concerning long scintillator strip, photons undergo many internal reflections on its way from the point of creation to photomultiplier. Number of reflections is dependent on the scintillator size and photon emission angle. Photon registration probability is influenced also by other factors e.g. absorption in the scintillator material, losses at the surface imperfections and quantum efficiency of the photomultiplier.

For the analysis of the light absorption in material, an effective absorption coefficient ($\mu_{eff}$) was calculated, scaling the absorption coefficient of pure polystyrene [71] to the experimental results obtained with the single detection unit of the J-PET detector (by factor of 1.8). The dependence of the $\mu_{eff}$ on wavelength is shown in Fig. 16.

Emission of photons with larger wavelength is profitable also considering light attenuation. In Fig. 16 emission spectra for the BC-420 and J-PET scintillators are compared to the effective light absorption coefficient (blue dotted line) $\mu_{eff}$ established for the BC-420 scintillator strips with rectangular cross section of 7 mm × 19 mm [70]. This value is not given by the producer. The coefficient was calculated relying on the absorption coefficient of pure polystyrene [71]. It is connected with self-absorption of light in scintillators. The light is being lost due to the transport of photons through the scintillator bar [70]. Absorption coefficient decreases with increasing wavelength. It indicates that the light attenuation (proportional to $e^{-\mu_{eff}}$) will be less for the J-PET scintillator with respect to the BC-420 since J-PET spectrum is extended towards larger wavelengths.

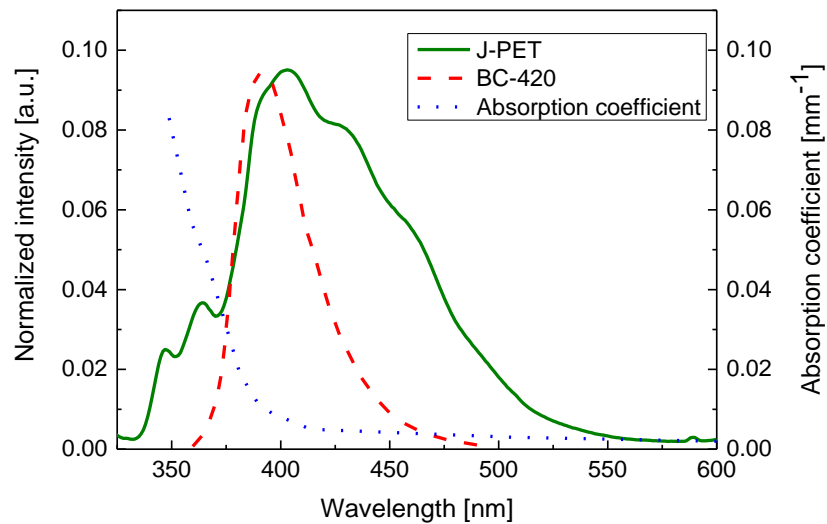

**Figure 16 Emission spectra of J-PET (green solid line), BC-420 scintillator (red dashed line) and absorption coefficient $\mu_{eff}$ (blue dotted line) [70] [71]. The emission spectra are normalized in amplitude.**

## 7.2. Spectral properties of J-PET scintillator

One of the most important parameters that characterizes scintillators performance is light output. It is defined as a number of emitted photons per unit of energy. An extremely effective scintillators are inorganic crystals e.g. CsI(Tl) - 65000 photons/MeV, NaI(Tl) - 42000 photons/MeV, $Lu_2SiO_5$:Ce (LSO) - 25000 photons/MeV, $BaF_2$ - 9500 photons/MeV and $Bi_4Ge_3O_{12}$ (BGO) - 8200 photons/MeV [72]. Organic crystals also can act effectively as scintillators, for example light output of anthracene is equal to 17400 photons/MeV [73].

In general, light output of plastic scintillators, which also belong to the group of organic scintillators, are much lower in comparison to the most effectively working crystalline scintillating materials. Light output of plastic scintillators produced by Saint Gobain range from about 30 % to about 80 % of the anthracene light output, for the majority of them it is about 60 % [12]. In this chapter the light output of the J-PET scintillator is examined.

Positron Emission Tomography examinations are based on registration of annihilation gamma quanta with energy of 511 keV. Due to that, light output of J-PET scintillators was determined by irradiating the scintillator with annihilation gamma quanta of this energy emitted by $^{22}$Na isotope. The beam of gamma quanta was collimated by collimator with 1.3 mm slit. The scheme of experimental setup which was used to determine the light output of J-PET scintillator is presented in Fig. 17.

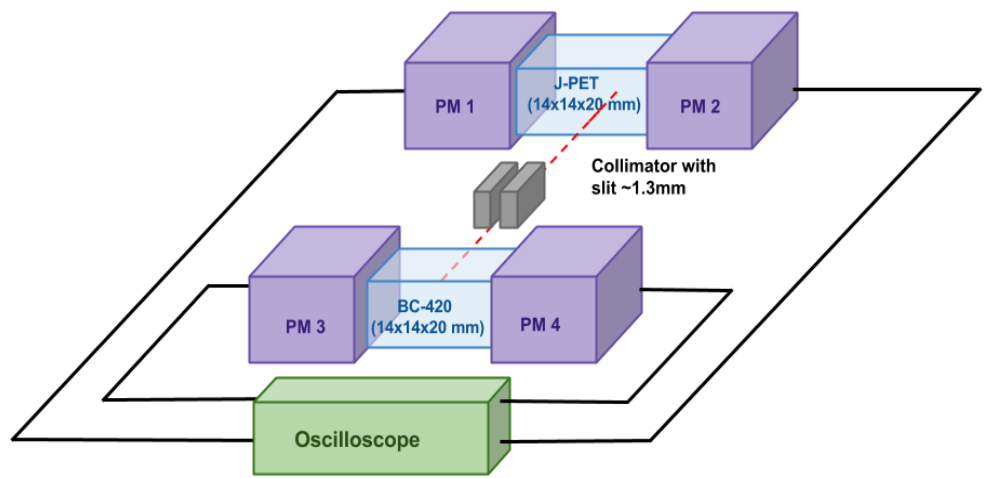

**Figure 17** A scheme of experimental setup used to determine J-PET scintillators light output. Performance of J-PET scintillator was compared to BC-420 scintillator. Detailed description of the setup and experimental procedure is given in the text.

In the experimental setup two scintillators are placed: J-PET and BC-420 with identical dimensions: 14 mm × 14 mm × 20 mm. Scintillator are wrapped in Vikuiti specular reflective foil [74] to prevent the loose of light. Each scintillator was connected by EJ-550 optical gel [13] to the window of R9800 Hamamatsu photomultiplier tube [21] at both ends. Electric signals coming from photomultipliers were sampled by the oscilloscope with 100 ps interval.

Light output of J-PET scintillators was determined with respect to known light output of BC-420 scintillator. That is why BC-420 was used as reference scintillator in the above experimental setup.

Plastic scintillator consist of low atomic number (Z) elements, mainly carbon and hydrogen. Gamma quanta interact with electrons of low Z elements predominantly via Compton effect [75]. Therefore the charge distribution is continuous. Photoelectric maximum, typical for crystal scintillator is not observed.

The charge of registered signals is proportional to the number of scintillation photons which in turn is proportional to the energy transferred by gamma quantum to an electron of scintillator. In our experiment, the energy of gamma quanta is fixed and equal to 511 keV, the maximum of possible energy transfer is also well defined and equal to 341 keV. Determination of the J-PET scintillators light output is possible by comparison of charge of signals for both: commercial and manufactured scintillators. Preliminary studies of the light output of the J-PET scintillator light output are subject of the article [76], and the detailed results described below are submitted for publication [77].

Measurements were carried out for a series of eight scintillators with different concentration of 2-(4-styrylphenyl)benzoxazole, varying from 0 to 0.5 wt. ‰. In order to minimize the influence of instrumental uncertainties coming from e.g. photomultipliers miscalibration, tests were performed twice for each sample, exchanging position of J-PET and BC-420 scintillator in the experimental setup. Thus, charge spectra registered by the same pair of photomultipliers can be compared.

Charge spectra registered for the series of J-PET scintillators containing different amounts of wavelength shifter, and BC-420 by two pairs of photomultipliers: PM1, PM2 and PM3, PM4 are presented in Fig. 18. Left panel shows spectra obtained with PM1 and PM2 photomultipliers and right panel presents spectra measured with PM3 and PM4. Spectra are arranged in order of ascending concentration of wavelength shifter.

Dependence of light output on the concentration of wavelength shifter in plastic scintillator was determined in order to set the optimal concentration of 2-(4-styrylphenyl)benzoxazole, which enables the most effective scintillator performance.

Comparison of the charge spectra registered by each photomultiplier were conducted using two methods. In the first one [10], scaling factor (LR) is calculated, by which the charge of J-PET scintillator needs to be divided to obtain spectra of J-PET and BC-420 fitting together. The scaling factor is equal to the ratio of particular scintillators light output:

$$LR = \frac{Q_{J-PET}}{Q_{BC-420}} \qquad (2).$$

Scaling factor was calculated for each photomultiplier. Values of light output given in the second column of Tab. 8 are averaged results obtained for four photomultipliers.

In the second method LR was calculated as a ratio of the middles of Compton edges which are the right edges of spectra. The ratio is interpreted as the relative light output. Knowing the light output of BC-420, light output of J-PET scintillators was calculated. The middle of the Compton edge of each spectrum was determined by fitting the Novosybirsk function to the edge of charge spectrum [78]. Values of J-PET scintillators light output determined using this method are given in the third column of Tab. 8.

Differences in the beginning of the spectra, around 0 pC are caused by fact that PM1 and PM3 were used in conditional trigger mode. Signals were registered by PM3 if a signal has been registered by PM1 within defined time window. Differences at the end of spectra, which is the Compton edge, are due to different light output of scintillators.

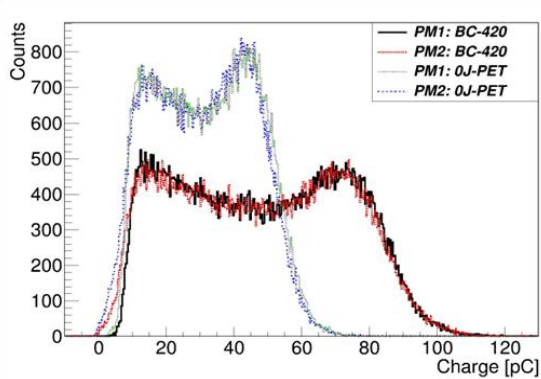
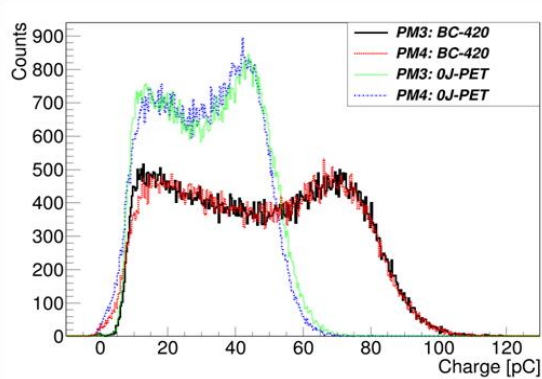
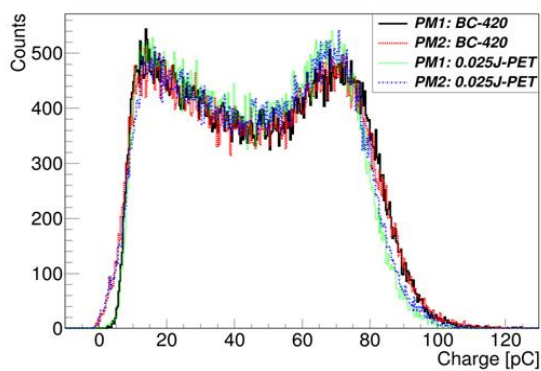
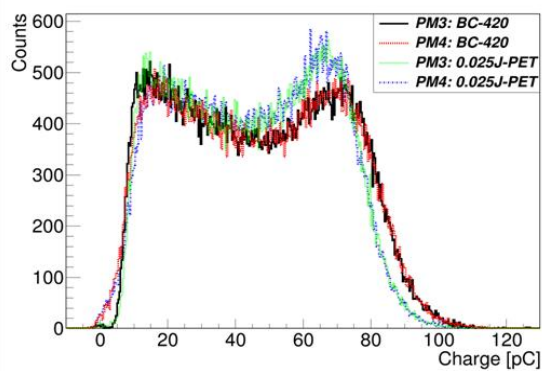
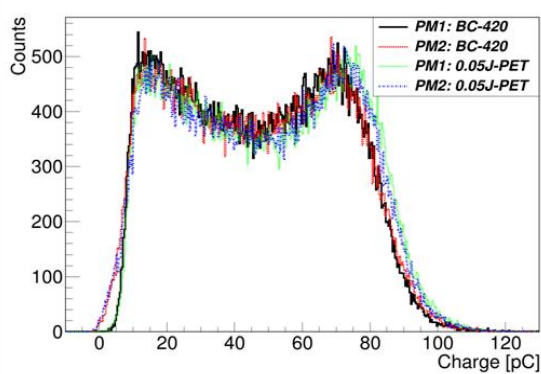
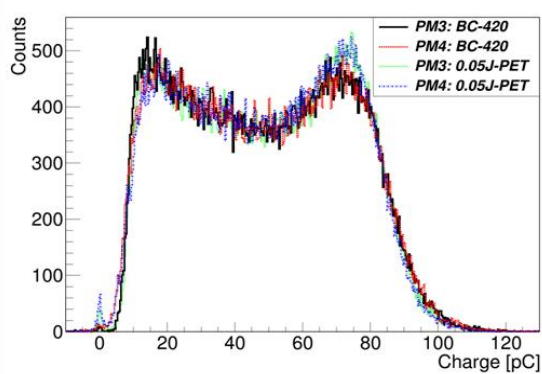
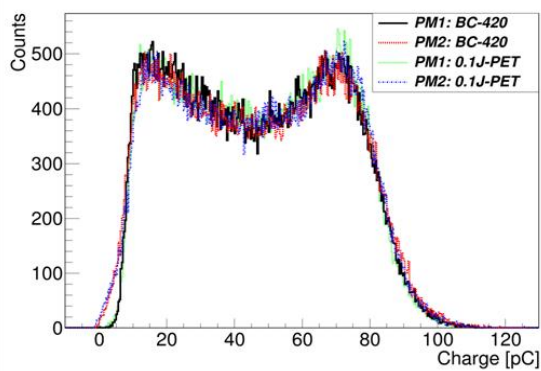
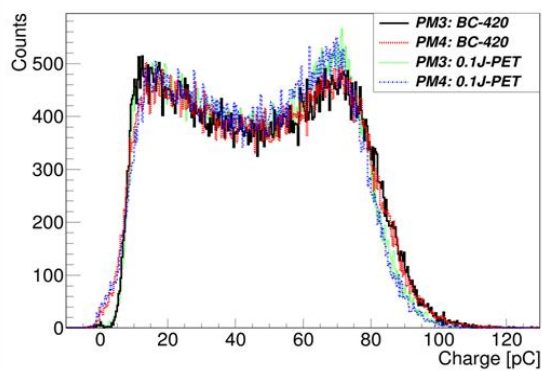

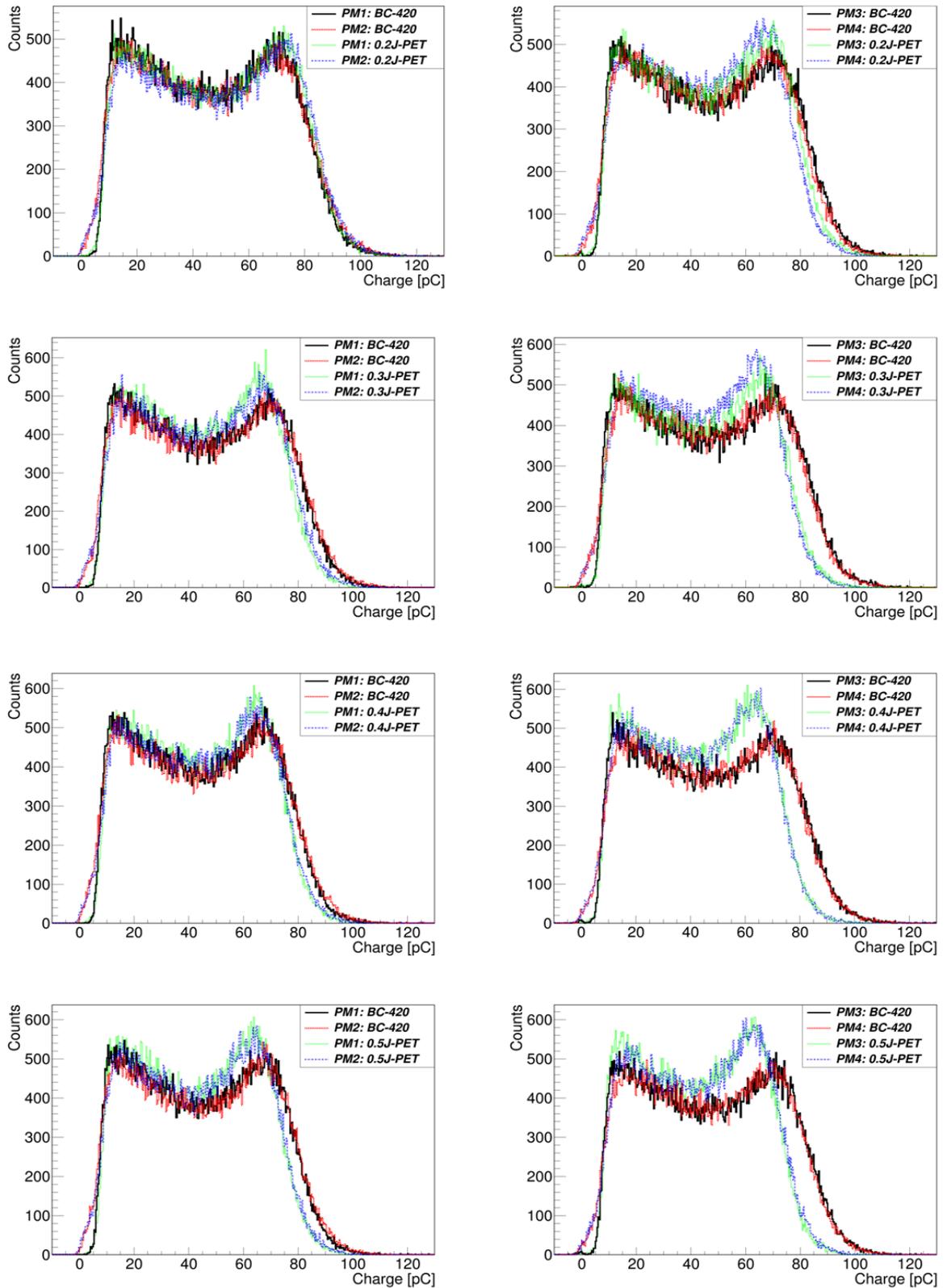

**Figure 18 Charge spectra registered for BC-420 and J-PET scintillators. PM1, PM2, PM3 and PM4 denote particular photomultipliers, according to Fig. 17. Spectra are registered for J-PET scintillators containing different amounts of novel wavelength shifter, from 0 to 0.5 ‰. Names of scintillators refer to the amount of wavelength shifter. In the left panel spectra registered by PM1 and PM2 for BC-420 and J-PET are presented, while in the right panel spectra registered by PM3 and PM4 are given.**

Table 8 Relative and absolute values of light output for J-PET scintillators determined for several concentrations of wavelength shifter. Relative light output was determined with respect to the BC-420 scintillator.

| WLS concentration [‰] | Method 1 | Method 2 | Relative light output | Light output [photons/MeV] |
|---|---|---|---|---|
| 0 | 0.636 ± 0.020 | 0.625 ± 0.002 | 0.625 ± 0.002 | 6397 ± 25 |
| 0.025 | 0.939 ± 0.008 | 0.938 ± 0.016 | 0.939 ± 0.007 | 9613 ± 72 |
| 0.05 | 1.009 ± 0.018 | 1.006 ± 0.021 | 1.008 ± 0.014 | 10318 ± 139 |
| 0.1 | 0.968 ± 0.150 | 0.967 ± 0.019 | 0.967 ± 0.019 | 9902 ± 197 |
| 0.2 | 0.965 ± 0.345 | 0.962 ± 0.031 | 0.962 ± 0.031 | 9853 ± 319 |
| 0.3 | 0.913 ± 0.017 | 0.912 ± 0.023 | 0.913 ± 0.014 | 9346 ± 142 |
| 0.4 | 0.910 ± 0.016 | 0.910 ± 0.035 | 0.910 ± 0.015 | 9317 ± 149 |
| 0.5 | 0.895 ± 0.023 | 0.897 ± 0.031 | 0.896 ± 0.018 | 9172 ± 188 |

Results obtained by method 1 and method 2 are consistent within error bars. Statistical uncertainties coming from fitting in both methods are of order of one-hundredth of the light output value. Therefore, they are negligible in comparison to the systematic uncertainties resulting from measurement technique with different photomultipliers.

Uncertainties of light output value given in the second and third columns of Tab. 8 were estimated conservatively as a half of a difference between the maximal and minimal values obtained from measurements with different photomultipliers.

In the fourth column of Tab. 8. results of the relative light output are presented. They were calculated as a weighted average light output values determined with method 1 and method 2.

In the fifth column of Tab. 8 results of the absolute light output determined for J-PET scintillators are given. The values were obtained by multiplying of the relative light yield by the known light output of BC-420 which is equal to 10240 photons/MeV [12]. The relative and absolute light output dependence on the wavelength shifter concentration in the J-PET plastic scintillator is presented in Fig. 19.

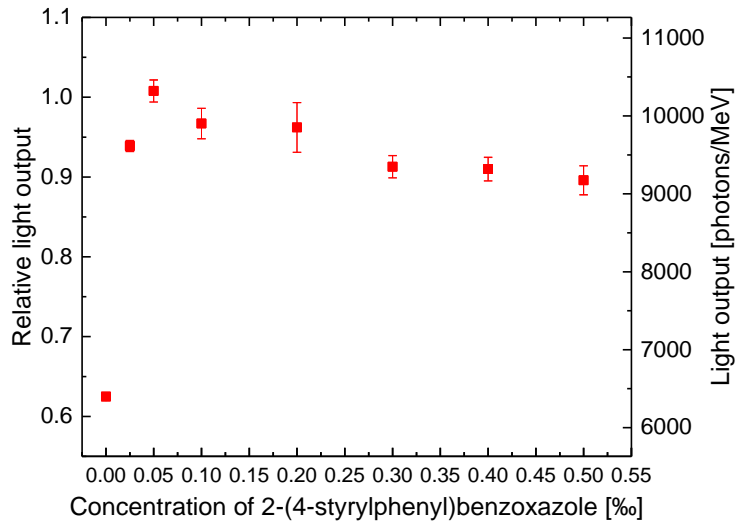

**Figure 19 Relative and absolute light output of J-PET plastic scintillators. The relative light output was determined with respect to the BC-420 plastic scintillator.**

Based on data presented in Fig. 19 and Tab. 8, one concludes that the highest light output is achieved for J-PET scintillator containing 0.05 ‰ of novel WLS, so called 0.05J-PET. It corresponds to 10318 ± 139 photons/MeV. These values meet the light output of the light output of BC-420, which is equal to 10240 photons/MeV.

The light output of J-PET scintillators of lower and larger concentrations of 2-(4-styrylphenyl)benzoxazole, is lower than in case of 0.05J-PET. Up to the concentration of 0.05 ‰, the amount of wavelength shifter is inefficient to enable an effective energy transfer. Considering values larger than 0.05 ‰, light loss is due to concentration quenching [79]. 0J-PET scintillator does not contain wavelength shifter, only polymeric matrix and primary fluor. Wavelength of the light produced by this sample lies within UV range which is not matched to the quantum efficiency of photomultipliers and is strongly absorbed in the material. That is why less light is detected and the measured light output of the scintillator has significantly lower value in comparison to other samples.

Optimization of wavelength shifter concentration was carried out for scintillators samples with dimensions 14 mm × 14 mm × 20 mm. However, according to the article [31], concentration of wavelength shifter in plastic scintillator is not a fixed characteristics. It is dependent on its length and even the shape. In larger J-PET scintillator strips, the optimal concentration of WLS will be probably lower than in tested 2 cm long scintillators. Results described in the thesis prove that 2-(4-styrylphenyl)benzoxazole can act as an effective wavelength shifter in plastic scintillator and its action is comparable with commercially used substances.

## 7.3. Timing properties of J-PET scintillator

In order to characterize signals appearing in the J-PET scintillator, rise and decay times were determined. Rise time was calculated as a difference between time in 10 % and 90 % of the signal amplitude on the leading edge: $t_{10\text{-}90}$. To analyze the shape of signals in scintillators, average signals registered by photomultipliers for 0.05J-PET and BC-420 were plotted (Fig. 20).

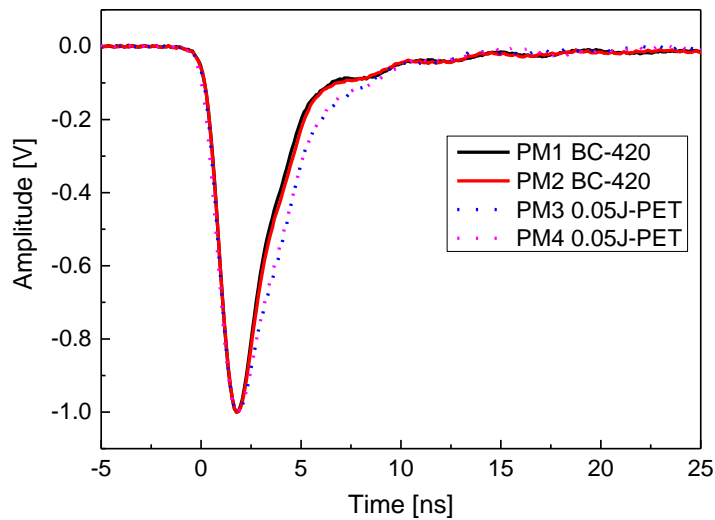

**Figure 20 Average signals appearing in 0.05J-PET and BC-420 scintillator, normalized to the amplitude of 1 V. PM denotes particular photomultiplier.**

The shape of signals shown in Fig. 20 is a convolution of the temporal density distribution of photons emitted by the scintillator, shape of the single photoelectron signal generated by photomultiplier, transit time spread (TTS) and the broadening of signal due to the finite bandwidth of the oscilloscope. Thus in general the rise time of the light signal ($T_{scintillator}$) may be extracted from the rise time of the measured electric signal ($T_{experiment}$) using a following formula (3):

$$T_{scintillator} = \sqrt{T_{experiment}^2 - T_{photomultiplier}^2 - T_{oscilloscope}^2 - TTS^2 - T_{eff}^2} \quad (3),$$

where $T_{photomultiplier}$, $T_{oscilloscope}$ and TTS denote the contribution to the observed rise time due to the form of the single photo-electron, oscilloscope bandwidth and transit time spread, respectively. $T_{eff}$ denotes the overall effective contributions due to the possible deviation of the nominal values of the above mentioned properties from the values provided by the manufacturers. $T_{photomultiplier}$ is equal to 1 ns [21], $T_{oscilloscope}$ calculated as a ratio 350/bandwidth [80], is equal to 0.07 ns while TTS value is 0.27 ns [21]. Mean rise time of the electric signals: $t_{10-90}$ is equal to $T_{experiment}$=1.22 ± 0.02 ns for BC-420 and 1.24 ± 0.02 ns for J-PET scintillator, respectively. This implies that within the measurement uncertainties the rise time of the scintillation in the J-PET scintillator is equal to the rise time of the BC-420 scintillator and amounts to 0.50 ns [12].

Although the rise time of 0.05J-PET and BC-420 scintillators have the same values, there is a slight difference in their decay time visible even in signals shapes in Fig. 20. Spectrum of 0.05J-PET is broaden in the region of the trailing edge in comparison to spectrum of BC-420 scintillator. This suggests that the difference in decay time values in both scintillators will be noticeable.

In ternary plastic scintillators the distribution of the time of photon emission followed by the interaction of the gamma quantum at time Θ, is given by formula 4 [81] [82]:

$$f(t|\Theta) = K \int_{\Theta}^{t} \left( e^{-\frac{t-\tau}{t_d}} - e^{-\frac{t-\tau}{t_r}} \right) \cdot e^{-\frac{(\tau-\Theta-2.5\sigma)^2}{2\sigma^2}} d\tau \quad (4).$$

The Gaussian term with the standard deviation σ reflects the rate of energy transfer to the primary solute, whereas $t_r$ and $t_d$ denote the average time of the energy transfer to the wavelength shifter and decay time of the final light emission, respectively. K stands for the normalization constant.

Decay time of signals in plastic scintillators was determined by fitting sum of functions given by formula 4 and formula 5 [83]:

$$f = \text{N}(sin^2(\alpha t + \varphi) + \Delta)f_{exp} \cdot f_{Gauss} \qquad (5),$$

where N, $\alpha$, $\varphi$ and $\Delta$ denote fitting parameters, $f_{exp}$ - exponent function and $f_{Gauss}$ denotes the Gauss function.

The formula consisting of the sum of functions given by formula 4 and 5 was fitted to averaged signals registered by particular photomultipliers. The fit is shown in Fig. 21.

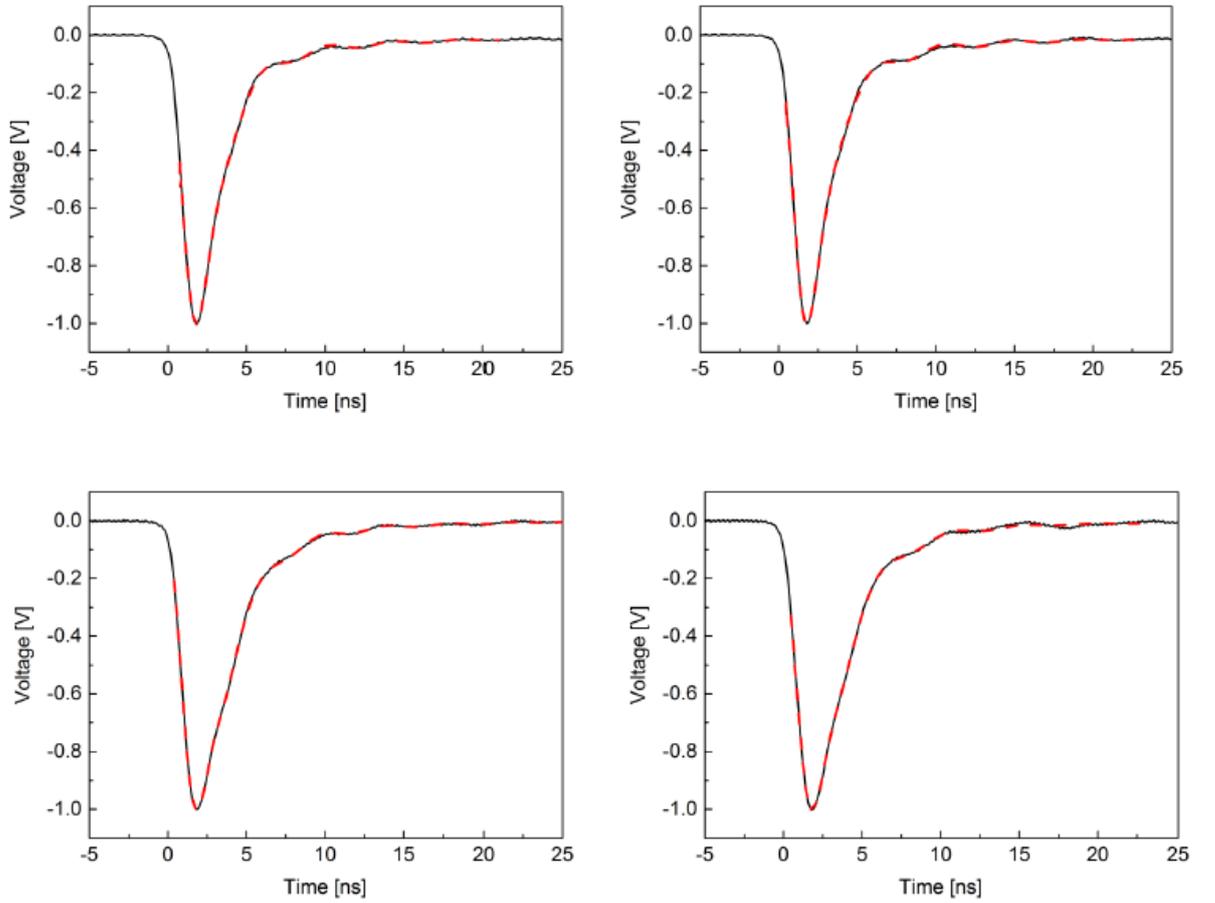

**Figure 21 Averaged signals from photomultiplier connected to BC-420 (top panel) and J-PET scintillator (bottom panel). Superimposed dashed-red lines indicate result of the fitting of sum of functions given by formulas 4 and 5. Signals were normalized to 1 V.**

Decay time of signals registered by photomultipliers connected to BC-420 scintillator determined by the method described above is equal to about $1.49 \pm 0.02$ ns. This is the value consistent with 1.5 ns, which is declared by the producer [12].

However, there is a discernible difference between decay time of signals in BC-420 and 0.05J-PET scintillator in which it is equal to $1.91 \pm 0.03$ ns.

Decay time is an important parameter characterizing plastic scintillators, especially these which will be applied in experiments based on measurements of signals time, like J-PET/MRI. It is one of the most important factors determining time resolution, which is a measure of the time interval in which two signals can be distinguished. The shorter the decay time is, the better is the scintillator.

BC-420 is one of the fastest plastic scintillator manufactured by Saint Gobain, what means that its decay time has very low value. However, decay time of signals in other plastic scintillators e.g. BC-400, BC-404, BC-408 are equal to 2.4 ns, 1.8 ns, and 2.1 ns, respectively. Therefore we can claim that decay time of signals in J-PET scintillator, which is approximately equal to 1.9 ns, is comparable with BC commercial scintillators and they are suitable for application in PET/MRI devices.

## 8. Structure of J-PET scintillators

### 8.1. Molecular weight

Condition of polymerization process and Trommsdorff effect influence the molecular weight of polymers. As it is apparent from article [84], there is a dependence of plastic scintillators light output on the polymer molecular weight (Fig. 22). Up to a value of about $10^5$ u, the light output increases with increasing polystyrene molecular weight. Above this value, light output reaches plateau.

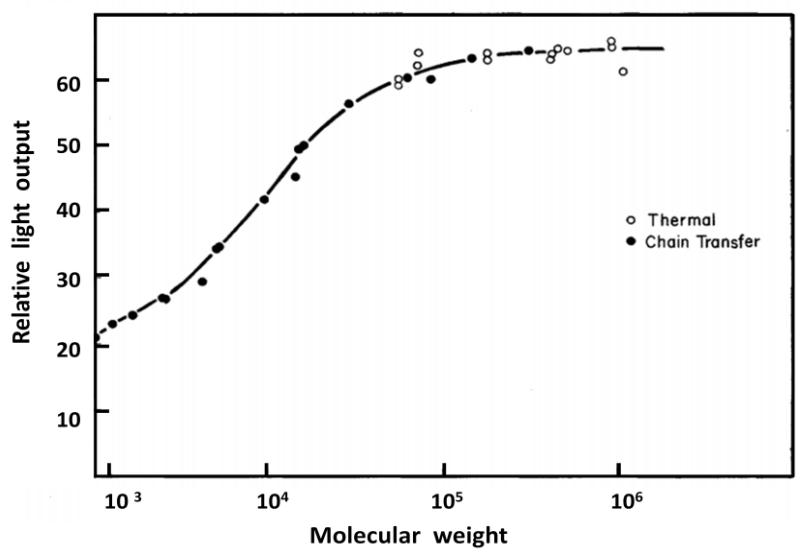

**Figure 22 The influence of molecular weight on the relative light output of plastic scintillators. Filled dots refer to samples in which molecular weight were changed by chain transfer agents, while empty dots refer to samples in which molecular weight was changed by polymerization temperature. The figure is adapted from [84].**

Because of the inhomogeneity of macromolecules molecular weight, it is necessary to calculate and determine its mean value. There are several methods of mean molecular weight description and their determination. In this thesis the viscous mean molecular weight was determined. This is one of the most popular methods of macromolecules molecular weights determination, which can be applied in the wide range of masses. This method is based on the rule that viscosity of polymers solutions is dependent on their molecular weight [85].

For the measurements the solution of polystyrene in toluene was prepared. Polystyrene was obtained in the same condition as J-PET plastic scintillators. Time of flow of solvent (toluene) and solutions with different concentrations of polystyrene between points A and B in Ostwald's viscometer (Fig. 23) were measured.

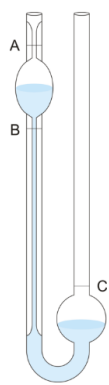

**Figure 23 Ostwald's viscometer. The picture is adapted from [86].**

Each time the viscometer was filled by the same amount of liquid by the right arm, to the point denoted as C. Then the liquid was suctioned above point A, and the time between reaching lines A and B was measured.

Then relative viscous ($\eta_r$) was determined for solvents of each concentration as:

$$\eta_r = \frac{\eta_p}{\eta_0} = \frac{t}{t_0} \qquad (6),$$

were $\eta_p$ is a viscosity of polymer solution, $\eta_0$ - viscosity of the pure solvent, $t$ - time of polymer solution flow between points A and B of the viscometer, $t_0$ - analogues time measured for the pure solvent. Specific viscosity was determined as:

$$\eta_{sp} = n_r - 1 \qquad (7),$$

and reduced viscosity is equal to:

$$\eta_{red} = \frac{\eta_{sp}}{c} \qquad (8),$$

where c denotes polymer concentration.

Based on the reduced viscosity dependence on the polymer concentration in the solution, the intrinsic viscosity [η] was determined by interpolating linear fit to the zero concentration. For data presented in Fig. 24, it is equal to 97.4.

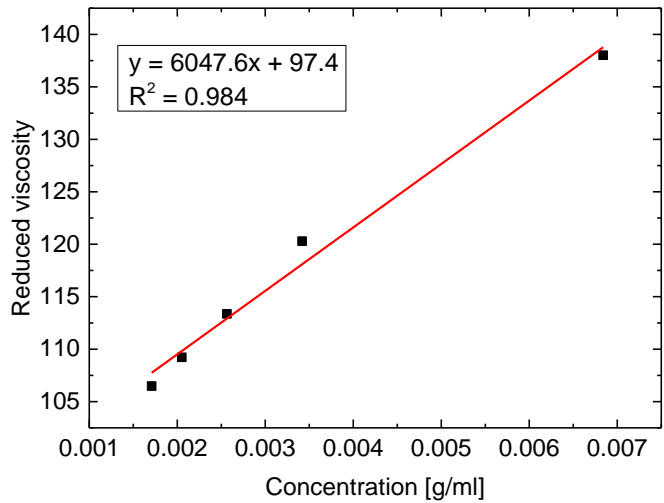

**Figure 24 Determination of intrinsic viscosity for a polystyrene solution in toluene.**

Basing on the Kuhn-Mark-Houwink-Sakurada equation:

$$[\eta] = K \cdot M_\eta^\alpha \quad (9)$$

molecular weight of the polymer (M) was determined. Viscometry is an indirect method of molecular weight determination and requires knowledge of constant values: α and K. For polystyrene solution in toluene measured in 25 °C, these constants are equal to [87]:

K= $10.5 \times 10^{-3}$ [ml/g]
α = 0.73.

Three samples of polystyrene obtained in different syntheses but the same conditions were tested using this method. Values of determined molecular weight are shown in Tab. 9.

Table 9 Molecular weight of polystyrene.

| Sample number | Molecular weight |
|---|---|
| 1 | $2.5 \times 10^5$ u |
| 2 | $2.8 \times 10^5$ u |
| 3 | $2.9 \times 10^5$ u |

The exact values of polymers molecular weights are not very important in this case. The most significant is that determined values of molecular weights are higher than $10^5$ u.

Since plastic scintillators are prepared in the same conditions, with identical temperature schedule, it is assumed that molecular masses of J-PET scintillators are equal to molecular weight of pure polystyrene. This means, that the light output of J-PET scintillators is maximized taking into account only polymer properties. Because polystyrene molecular weight is more than 2 times larger than $10^5$ u, light output of the scintillators is not sensitive to small variations of this quantity.

J-PET scintillators are based mainly on polyvinyltoluene, not polystyrene. However constant values: K and α are not determined for polyvinyltoluene solutions, so we took into consideration only results for polystyrene. We assume that polyvinyltoluene as polystyrene homolog, differing only by an additional methyl group in the mer, polymerizes in the same way and its molecular weight is similar.

## 8.2. Investigations with Positron Annihilation Lifetime Spectroscopy

Positron Annihilation Lifetime Spectroscopy (PALS) is an extremely valuable technique of materials structure characterization. It is especially useful in polymers study, since in vast majority of them positronium is formed and trapped. Measurements of positronium mean lifetime enable estimation of free volumes sizes in the structure. Characteristic properties of materials based on polymers are dependent on their structure which is determined inter alia by unoccupied regions which has an access to segmental motions. Ratio and sizes of free volume in the polymer have an impact on the mechanical properties and ageing of the material.

In this chapter structure of plastic scintillators will be described by determining temperatures of structural changes in the material. The most important is glass transition temperature ($T_g$) of amorphous polymers. It indicates the point in which properties of polymer are changed from the rigid glassy solid to more flexible state [88]. At temperatures below $T_g$ polymeric chains do not move around. After delivering thermal energy, motion is allowed. Macromolecules move around each other and amorphous rigid structure starts to be changed to the flexible one. It is connected with the change of heat capacity. Because of that $T_g$ can be determined by widely used methods of polymer structure characterization, e.g. Differential Scanning Calorimetry (DSC). However, it is also possible to determine temperatures of glass and other structural transition using PALS, basing on ortho-positronium mean lifetime and the intensity of its production.

In this method, positron is emitted from the radioactive source and travels a distance in the examined material, what is connected with the loose of its kinetic energy. When slowed down, positron can annihilate in two ways. The first one is free annihilation, with an electron encountered in the material. The second possible kind of positron annihilation is forming a meta-stable quantum mechanical state called positronium. It is a hydrogen-like atom consisting of positron and electron.

Depending on the positron and electron spin orientation, para- and ortho-positronium can be formed. In para-positronium (p-Ps) spins are orientated antiparallel. Lifetime of p-Ps is equal to 0.125 ns. In ortho-positronium (o-Ps) spins are parallel and its lifetime is much longer, equal to 142 ns in vacuum. However in condensed matter the lifetime can decrease even to 1-5 ns, because positron from o-Ps annihilates with an electron from surroundings having an opposite spin. That leads to two gamma ray annihilation and it is called the pick-off annihilation [89].

PALS enables to detect pores with sizes from 0.1 nm to 100 nm. Free volumes in the matter are assumed as potential well, finite in depth, in which positronium atom annihilates [90] [91]. Formula (10) describes the relationship between o-Ps lifetime in the trap ($\tau_{po}$) and the radius of the free volume (R) approximated using Tao-Eldrup model with an infinite potential well:

$$\tau_{po} = 0.5 \times \left(1 - \frac{R}{R+\Delta R} + \frac{1}{2\pi} sin \frac{2\pi R}{R+\Delta R}\right)^{-1} \quad (10),$$

with ΔR - empirical parameter corresponding to overlapping of o-Ps wave function with surroundings, for polymers equal to 0.166 nm [89].

In the large number of materials free volumes are not perfectly spherical and modification in the described model are necessary [92] [93]. However, in polymers free volumes are not perfectly defined and they can be interconnected, so Tao-Eldrup formula was applied in the original form.

The structure of J-PET scintillators was studied by means of Positron Annihilation Lifetime Spectroscopy [94]. Measurements were conducted in laboratory of Institute of Physics, Maria Curie - Skłodowska University in Lublin with delay coincidence "fast - slow" spectrometer. The setup for such measurements was prepared as follows. The [22]Na source of the activity of about 0.48 MBq was placed between two slices of plastic scintillators as shown in Fig. 25. Such "sandwich" was arranged in the chamber (Fig. 26).

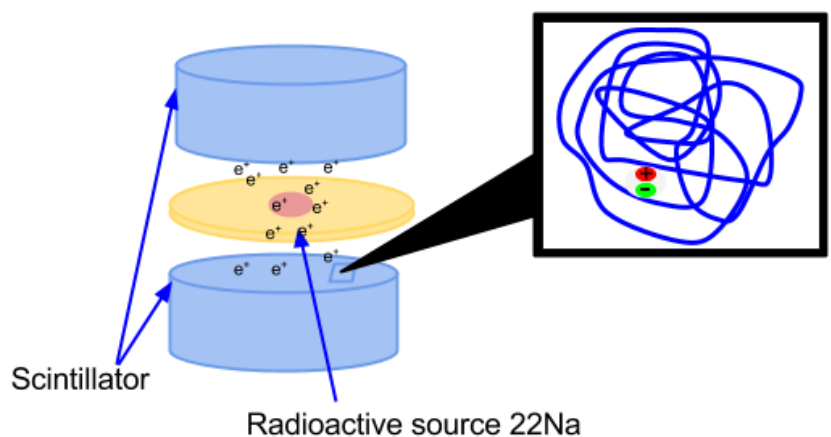

**Figure 25** The scheme of plastic scintillators samples arrangements in the experimental setup. Positrons emitted from sodium source interact with scintillating material forming positronium atom in empty voids between and around polymeric chains.

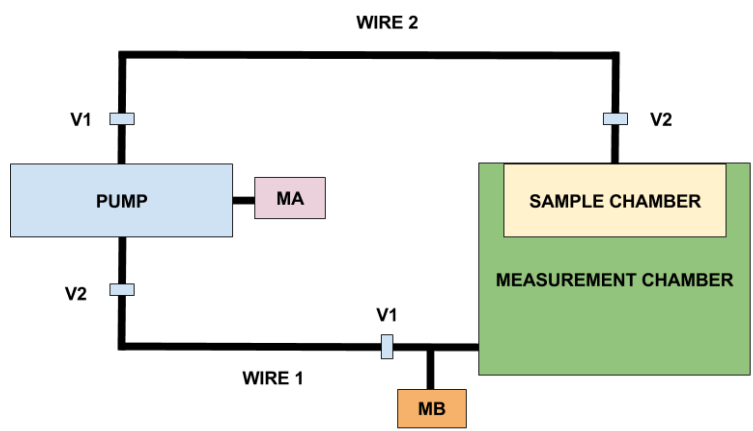

**Figure 26** Block scheme of the vacuum part of the experimental setup for positronium lifetime measurements. V1 and V2 denote valves, MA and MB denote measurers of vacuum.

The sample and source of positrons are placed in sample chamber enabling stabilization and control of vacuum and temperature. The apparatus shown in Fig. 26 consists of pump with measurers and system controlling supply valves. The turbomolecular pump enables obtaining vacuum of $2.1 \times 10^{-3}$ Torr.

Sample chamber is made of copper. In the bottom of chamber thermocouple is placed and below the heater is situated. It is wound on the copper rod which can be immersed in liquid nitrogen. Such setup with heater power supply connected with thermoregulatory device Shimaden FP 21 enables an automatic temperature control.

In measurements of positronium lifetime START signal corresponding to the time of the atom origin is regarded as registration of gamma quanta coming from $^{22}$Na source with energy of 1274 keV (Fig. 27). On the other detector STOP signal is registered, which is detection of gamma quanta of energy 511 keV. This is the energy of gamma quanta coming from annihilation. Therefore, positronium lifetime is the time interval between START and STOP signal.

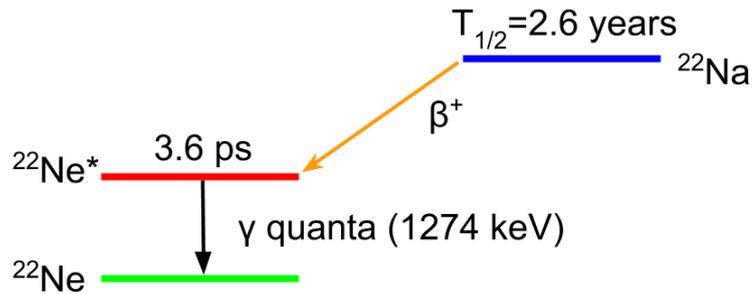

**Figure 27** The scheme of $^{22}$Na source decay in which gamma quanta of energy 1274 keV are produced just after emission of positron. Gamma quanta production is preceded by creation of non stable excited state of $^{22}$Ne.

Measurements were conducted with delay coincidence spectrometer which block scheme is shown in Fig. 28. Gamma quanta originating from tested sample reach scintillating detectors START and STOP. As scintillators $BaF_2$ crystals were used. Detectors emit scintillation of two components: fast (decay time shorter than 1 ns and wavelength 220 nm) and slow (decay time 620 ns and wavelength 310 nm). Detectors are characterized by large absorption of gamma quanta by photoelectric effect [95]. As photoelectric converters XP2020Q photomultipliers were used.

Spectrometer consists of two branches: time - which is fast, and energy - which is slow. Start and stop signals, corresponding to registration of quanta of energy 1274 keV and 511 keV are chosen by the energy branch. The role of the time branch is the measurement of time interval between registration of both quanta. Branches are conjugated therefore only coincidences fulfilling the condition of the energy selection are registered.

In the energy - "slow" - branch, the pulse coming from photomultipliers is amplified and shaped by spectrometric amplifier (SA). Next the pulse is transmitted to single channel amplitude analyzer (SCA). Signals with amplitude fitting to the fixed window of SCA are proceeded to triple coincidence system (TCS).

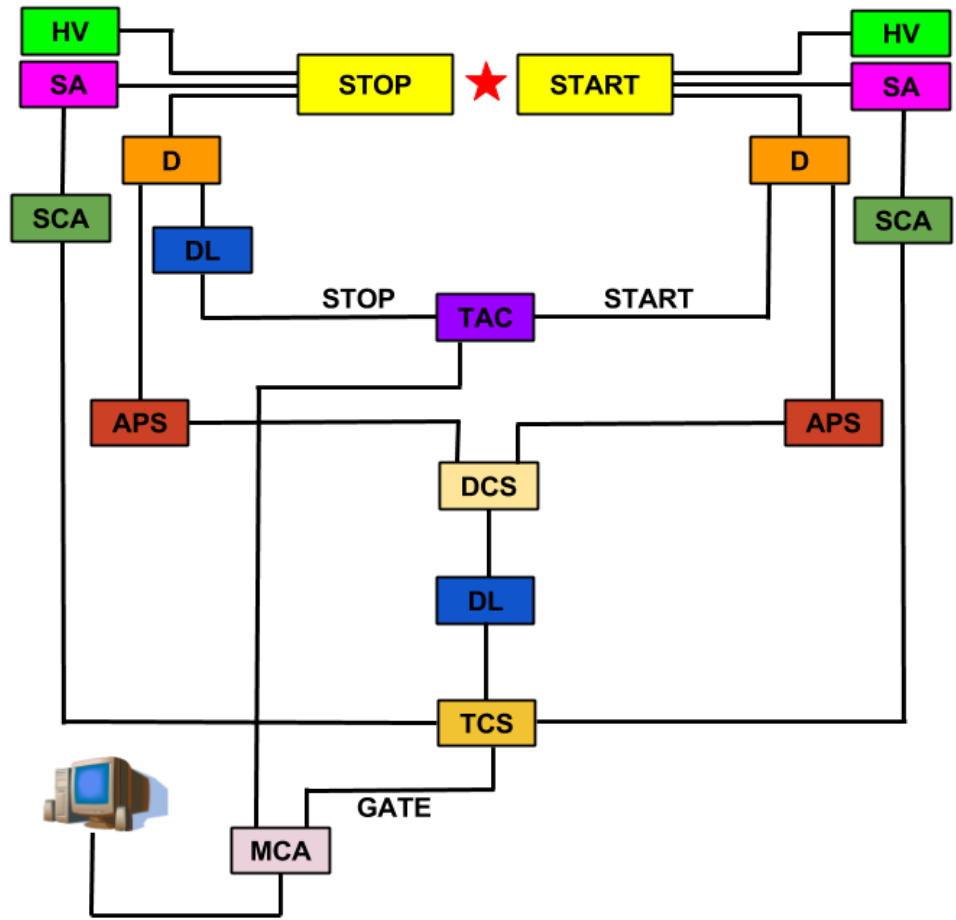

START - detector registering gamma quanta with energy of 1274 keV

STOP - detector registering gamma quanta with energy of 511 keV

HV - high voltage power supply

SA - spectrometric amplifier

SCA - single - channel amplitude analyzer

D - discriminator

TAC - time - to - amplitude converter

APS - anti pile - up system

DCS - double coincidence system

TCS - triple coincidence system

**Figure 28 Scheme of fast - slow spectrometer for measurements of positron lifetimes.**

In the time - "fast" branch, the pulse from photomultipliers is shaped by discriminator (D), which compensates time dispersion correlated with amplitudes and rise time of input pulse. From the discriminator two pulses are send. The first one is transmitted to anti pile up system, where overlapping pulses are eliminated. The second pulse reaches time - to amplitude converter (TAC). From the START detector the pulse is transmitted to TAC directly but from STOP detector through the delay line (DL). TAC generates a pulse of amplitude proportional to the time interval between START and STOP pulses. Then the pulse is transferred to multi channel amplitude analyzer (MCA) equipped with analogue - digital 8192 channels Wilkinson converter. Pulses from anti pile up system go to MCA as well through double coincidence system (DCS), delay line (DL) and triple coincidence system (TCS) which is a gate opening the entrance for the pulse coming from time - to amplitude converter (TAC). Data collected by multi channel amplitude analyzer may be read by computer.

Measurements were conducted within the temperature range from 123 K to 423 K. A temperature was raised using a resistance heater and liquid nitrogen was used in order to cool down the sample. Spectra were collected in each temperature for at least 2 hours. They were analyzed using LT 9.1 program [96]. Three discrete lifetime components were found: 170 - 190 ps which corresponds to p-Ps annihilation, 380 - 570 ps corresponding to free annihilation and one component over 1.8 ns which corresponds to o-Ps annihilation. The obtained results are presented in Fig. 29 showing o-Ps lifetime ($\tau_3$) and intensity ($I_3$) as a function of temperature. The o-Ps lifetime and intensity give information about size of free volumes and their concentrations, respectively.

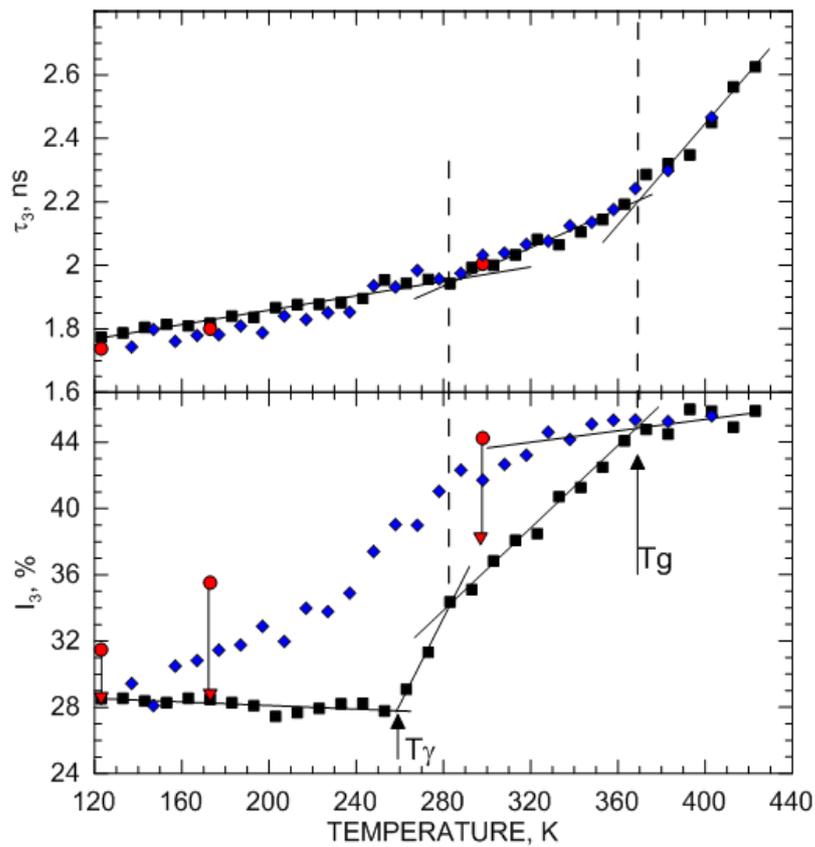

**Figure 29** Temperature dependence of $I_3$ in J-PET plastic scintillator. Points denote: squares - increasing temperature; diamonds - decreasing temperature; dots and triangles - first and last points in the measurements of irradiation effect; dashed lines and arrows denote the transition temperatures; red dots with arrows denote sections of time relaxation.

The ortho-positronium intensity is dependent on many factors, e.g. material purity, thermal history, the direction of temperature change, the change rate and the time of irradiation with positrons. The o-Ps lifetime ($\tau_3$) increases linearly with the temperature from 123 to 260 K within the range 1.75 ns to 1.95 ns. It stays at the same level up to 283 K, which is the point from which the growth rate of $\tau_3$ increases. The next growth rate increase starts in temperature 370 K. Both points in which $\tau_3$ changes its growth rate are correlated with changes of o-Ps intensity too.

Obtained results show that the o-Ps production intensity reveals a significant hysteresis connected mainly to matrix material. Moreover, three different rates of $\tau_3$ growth with increasing temperature (120 - 283 K, 283 - 370 K and 370 - 423 K) may indicate existence of three phases in tested scintillator. According to the article [97] where PALS investigations of pure polystyrene are described, glass transition in the scintillator occurs in 360 K. The temperature of glass transition of the tested J-PET scintillator is slightly shifted compared to pure polystyrene what may be caused by the presence of scintillating dopants. However, the range of o-Ps lifetime is similar to the results described in article [97].

Other structural changes, visible on the diagram of o-Ps intensity $I_3$ (bottom panel of Fig. 29), in 280 K and 260 K may be also correlated with structural changes in the doped polymer. The probable reason of the changes is crystalline - like organization of molecules in some regions of the amorphous polymer. One can find in the literature [98] information about phase transitions in polystyrene at low temperatures which is so-called beta-transition (between 283 K and 333 K). This point can be identified with the phase transition point found in our results in J-PET polystyrene - based scintillator (in 280 K). Red points in Fig. 24 denote rapid changes of $I_3$ dependence on time when the sample was stored in particular temperature. The measurement confirmed that o-Ps production is unstable in time (intensity $I_3$ decreases with time).

Differences between results obtained in our experiment and the experiment described in [97] come from the presence of dopants. It is known that even small amount of impurities may significantly change temperatures of particular phase transitions. For example in polypropylene copolymers and blends described in article [99] one can observe the shift of glass transition in comparison to sample of pure polystyrene. Because of the presence of admixtures, several regions of crystalline-like organization may be formed. It results in additional structural transitions. Transition denoted as $T_\beta$ is probably connected with the presence of dopants.

Considering the transition $T_\gamma$, it is clearly visible in the $I_3$ diagram but almost invisible in diagram of $\tau_3$. This indicates that the transition is correlated with energetic changes in the molecule, not a geometrical reconfiguration.

It was mentioned, that $I_3$ of o-Ps can be influenced by the thermal history of the sample. It is visible in the hysteresis observed for heated and cooled down scintillator. Because of that, an additional measurements of thermal stability were carried out. PAL spectra were collected for a long time, equal to at least 15 hours, in three temperatures: 123 K, 173 K and 298 K. Results of the measurements are shown in Fig. 30.

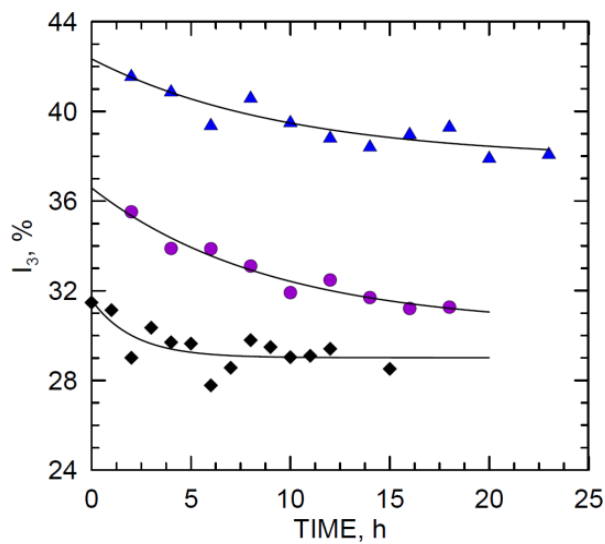

Figure 30 The ortho-positronium intensity (I₃) dependence on time in temperatures: 123 K (black diamonds), 173 K (violet dots) and 298 K (blue triangles) in scintillator based on polystyrene, containing 2 % of PPO and 0.03 % of 2-(4-styrylphenyl)benzoxazole.

In the room temperature (298 K) $I_3$ has relatively high value, equal to about 42 %. Then the sample was cooled down to 123 K and stored in this temperature for 15 hours. During this time, o-Ps lifetime ($\tau_3$) was stable (Fig. 29) but its intensity decreased significantly. Also in 173 K o-Ps production intensity is unstable in time. Time constants are equal to 2 h in 123 K, 8 h in 173 K and 10 h in 298 K. Time constants determine an amount of time which is needed to stabilize $I_3$ intensity. It is shorter for lower temperatures. In the article [100] in modified polystyrene similar effects were observed. Described time constants vary with temperature in range from 3 h to 30 h.

Fast decrease of intensity with time may be caused by competition between two processes: creation and decay of free radicals [100]. In case of our research, the o-Ps intensity in polymeric material is affected also by the temperature difference between the last point before jump and the temperature of long measurement. In the article [97] similar research were carried out, however the increase of $I_3$ in low temperature were observed, contrary to our research. The differences are caused probably by different polymeric material.

The size of free volumes was calculated assuming that they are spherical and using the Formula (4), and the results of these calculations are shown in Fig. 31. In the room temperature in which scintillators will be used, the volume of free voids is about 0.1 nm³. With increasing temperature, sizes of pores increase. At temperature of 300 K the size of free volume is doubled. In calculations it was assumed that volumes are spherical. However, in plastic empty voids can have different shapes, so obtained results are indicative and averaged over voids of different sizes and shapes.

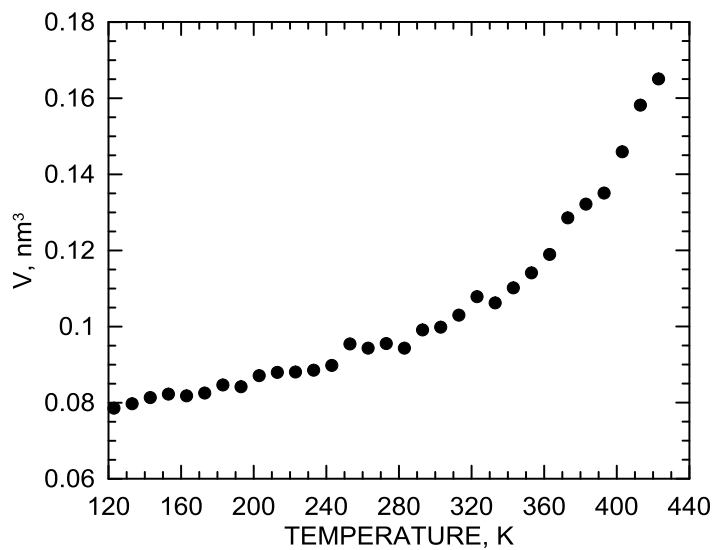

**Figure 31 The size of spherical free volume as a function of temperature in scintillator based on polystyrene, containing 2 % of PPO and 0.03 % of 2-(4-styrylphenyl)benzoxazole.**

Research carried out by PALS method, which are described above are the subject of article [94] by the author of the thesis and collaborators.

Because chemical structures of polystyrene and polyvinyltoluene are similar, differing only by the one methyl group connected to aromatic ring in mer, no large differences between PALS spectra of scintillators based on PS and PVT are expected. Samples of J-PET scintillator based on polyvinyltoluene and pure polyvinyltoluene were subjected to PALS measurements. J-PET scintillator beside polyvinyltoluene consists of 2 % of PPO and 0.05 ‰ of the novel WLS. This constitution is set for the most effective J-PET scintillator. In order to determine the influence of fluorescent additives on the structural transitions in scintillator, a sample of pure polyvinyltoluene was tested as well. Ortho-positronium lifetime ($\tau_3$) and intensity ($I_3$) were registered as a function of temperature. Conditions of measurements and analysis of obtained results were carried out in the way which is described previously, in the experiment which results are presented in Figs. 29 - 31. PALS results determined for J-PET scintillator based on polyvinyltoluene and for pure polyvinyltoluene are presented in Fig. 32.

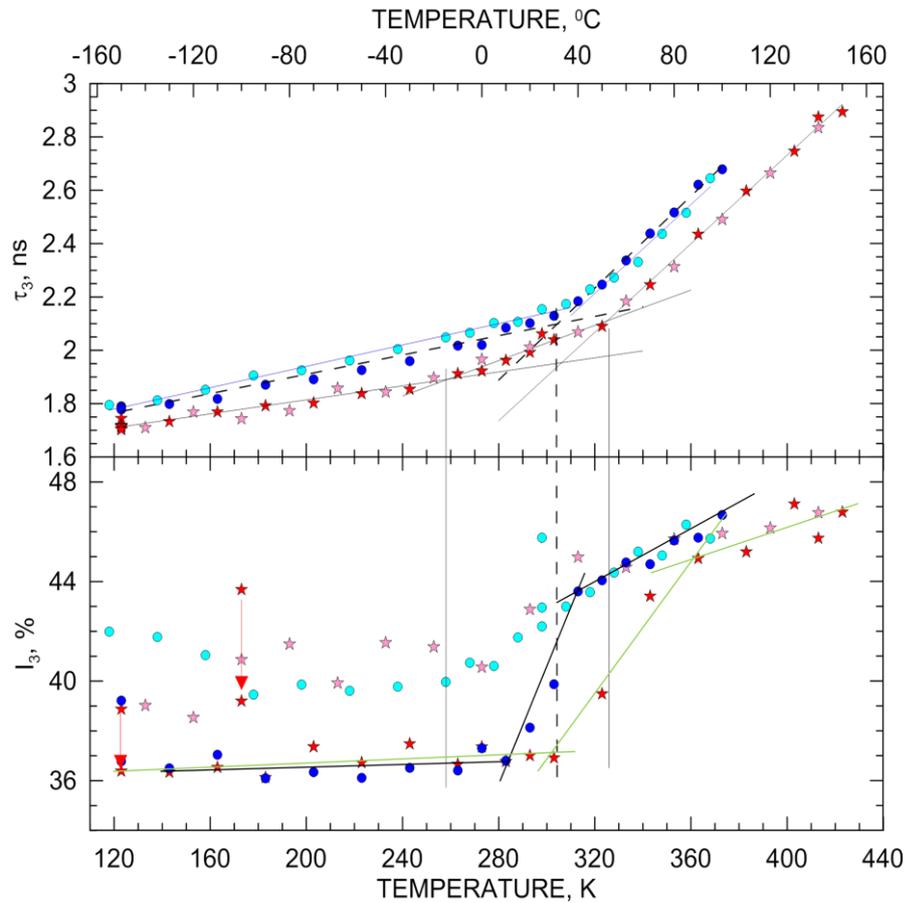

**Figure 32** Temperature dependence of ortho-positronium lifetime ($\tau_3$) and intensity ($I_3$) in J-PET scintillator based on polyvinyltoluene (stars) and pure polyvinyltoluene (dots). Red stars refer to J-PET scintillator when heating, pink ones, when cooling. Dark blue dots refer to pure polyvinyltoluene when heating, light blue dots, when cooling down.

In both: scintillator and polyvinyltoluene intensity of o-Ps production increases with temperature, however values obtained in particular temperatures during heating and cooling are different. Such hysteresis is observed also for J-PET scintillator based on polystyrene (Fig. 29). Three different rates of $I_3$ growth with increasing temperature are observed, for scintillator: 123 - 305 K, 305 - 360 K, 360 - 423 K and for PVT: 123 - 280 K, 280 - 310 K, 310 - 423 K. Points denoted temperatures growth changes for scintillator are slightly shifted towards larger temperatures in comparison to pure matrix.

Considering o-Ps lifetime in both samples, one can notice significant difference. There are three rates of $\tau_3$ growth registered for scintillator: 123 K - 260 K, 260 K - 325 K, 325 K - 423 K and only two for pure polyvinyltoluene: 123 K - 310 K and 310 K - 423 K. Existence of an additional point of $\tau_3$ growth change in the scintillator sample is connected with the presence of the additives.

Comparing $I_3$ and $\tau_3$ spectra of polymeric matrix, points indicating structural changes are placed within the similar temperatures, 305 K and 310 K, respectively. Considering results obtained for the scintillator sample, discrepancy in characteristic temperatures indicated by $I_3$ and $\tau_3$ can be observed. The first one in $I_3$ is positioned around 260 K, while in $\tau_3$ around 300 K. The second one, in $I_3$ is placed around 320 K, while in $\tau_3$ within 360 K. It may indicate that $I_3$ and $\tau_3$ are not equally sensitive for the structural change. It is probable that the change starts in 260 K or further in 320 K for particular polymeric change however in the macromolecule as a whole can be observed only in higher temperatures. Such result can be interpreted as an interval in which the change takes place.

The first temperature region is identified as $T_\gamma$ described in the caption of Fig. 29. The second characteristic temperature in which the rate of $I_3$ growth changes is around 305 K - 310 K for undoped polyvinyltoluene and 320 - 360 K for the scintillator. These are temperatures of glass transitions. Temperatures determined for scintillator and pure matrix differ significantly. Such differences is caused by the dopants which may create regions of local ordering. Glass transition of the polyvinyltoluene - based scintillator (360 K) is similar to the $T_g$ of polystyrene - based scintillator (370 K).

Tests of $I_3$ stability within time (as indicated by red arrows) show similar results to the sample of polystyrene-based scintillator. Time constants are shorter in low temperatures: few hours in 123 K and about ten in 173 K.

To summarize, results of PALS measurements conducted for polyvinyltoluene - based J-PET scintillator and the pure matrix, are shown in Tab. 10.

**Table 10 Temperatures of structural changes determined for J-PET scintillator and pure matrix determined by PALS.**

| Sample | $T_\gamma$ [K] | $T_\gamma$ [ºC] | $T_g$ [K] | $T_g$ [ºC] |
|---|---|---|---|---|
| PVT-based J-PET scintillator | 260 - 300 | -13 - 27 | 320 - 360 | 47 - 87 |
| Pure PVT | - | - | 305 - 310 | 32 - 37 |

## 8.3 Investigations with Differential Scanning Calorimetry

The method of materials structure characterization which is complementary to PALS but more popular and conventional is Differential Scanning Calorimetry (DSC). It is based on the influence of the temperature on physical properties of the sample. During the measurement, the differences of temperatures are measured between the tested and the reference sample, in which no structural transitions occur. Structural changes in the examined sample are correlated with changes of heat capacity. Relying on this, exo - and endothermic processes in the sample can be identified.

DSC measurements of polyvinyltoluene - based J-PET scintillator and pure polyvinyltoluene were carried out. Registered thermograms are presented in Figs. 33 and 34.

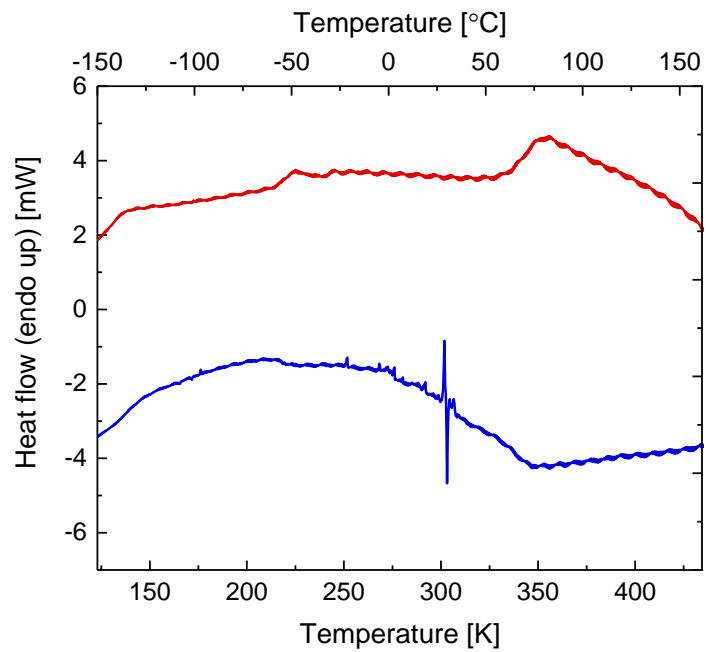

**Figure 33 Thermogram of J-PET scintillator containing 0.05 ‰ of novel WLS. Red line denotes heating, blue one cooling of the sample.**

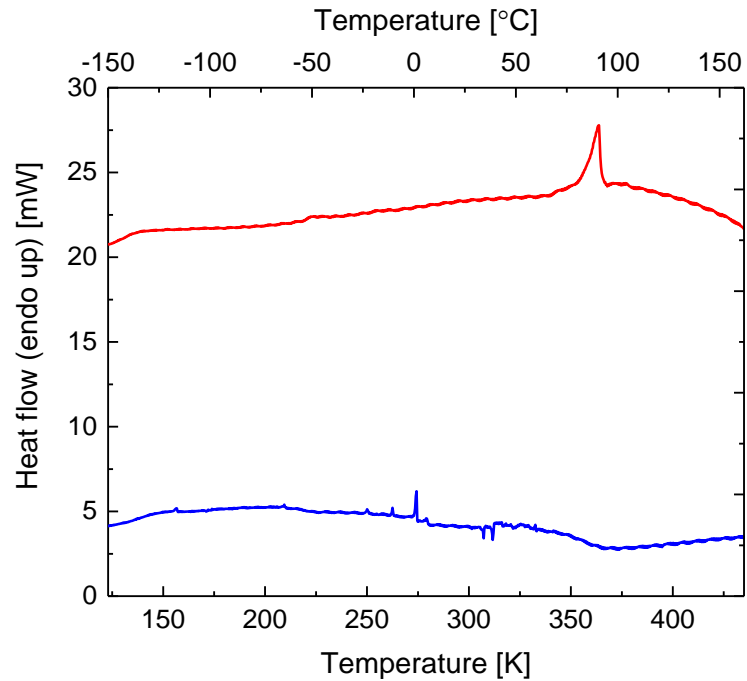

**Figure 34 Thermogram of pure polyvinyltoluene. Red line denotes heating, blue one cooling of the sample.**

In both cases around 350 K there are peaks corresponding to the glass transition temperature $T_g$. In the scintillator, glass transition determined from the heating curve, occurs in temperature 354 K while in pure polyvinyltoluene in 364 K. $T_g$ values determined using DSC are more similar to each other than results obtained from PALS analysis. They are also comparable with literature data. Glass transition temperature of polyvinyltoluene, depending on its structure, lies in range 366 - 391 K [101]. The difference of 2 K between the literature and experimental data is negligible because results are dependent on many factors, including the rate of heating.

In general, $T_g$ obtained in DSC are significantly larger in comparison to results of PALS. It is due to the fact that $T_g$ is associated with the onset of a coordinated segmented motion about polymer axis. The pick-off process of o-Ps is much more sensitive to such changes and reorganization in the macromolecule than DSC, in which only macroscopic effects can be observed, occurring in the sample as a whole [102]. Therefore temperatures determined by PALS may be interpreted as the onset of changes in the structure of macromolecule.

The other issue that should be discussed is time scale of both experiments. In PALS the sample was stored in each temperature for 2 hours, while in DSC the rate was set to 10 K/min. Different rate of temperature changes plays an important role even using the same technique and it has a discernible impact on obtained results [102]. Obtaining such discrepancy in results when using two different methods with two radically different rates of heating was expected.

Generally it is assumed that the softening point of amorphous polymers is 20 K below the glass transition point [103]. Therefore, softening temperature of J-PET scinillator is at 334 K (61 ºC). The value of softening point of commercial scintillators produced by Saint Gobain, which are also based on polyvinyltoluene is 343 K (70 ºC) [12]. Discrepancy between both values is relatively large, but the method of determination of softening point used by Saint Gobain is not given, so they cannot be compared with the high precision. Softening point can be interpreted as the maximal temperature in which scintillator can be used. However, PALS analysis indicates that changes in the J-PET scintillator structure start to occur in much lower temperature region including also room temperature in which scintillators are being used.

## 9. Status of production of J-PET plastic scintillator strips

Because results of tests performed with small scintillator samples are satisfactory, the next step was to manufacture plastic scintillator strips with much larger dimensions. For this purpose, special furnace (DCF420/spec) was designed to enable producing long strips. Scintillators described in the thesis so far, have been produced using furnace of CZYLOK [104] which enabled synthesis of only small size samples of length less than 5 cm because of geometry and size of the chamber.

Small scintillator samples produced previously, were polymerized in the glass ampoule placed vertically in the furnace, as the chamber is in the shape of pipe with diameter of few centimeters. Glass is a very good and cheap material for the forms to bulk polymerization of polystyrene and polyvinyltoluene because polymer does not adhere to the surface of form. Additional advantage of the glass form is perfect tightness when closed in the flame of the burner. Having finished polymerization process, the glass form (ampoule) can be broken and the material can be taken out easily.

However it is not possible to obtain longer scintillator in respectively longer glass ampoule placed in pipe furnace (left panel of Fig. 35). During polymerization in such form, probably the pressure is too high and empty voids (bubbles) are generated in the whole volume of the sample. What is more, scintillators manufactured in such ampoules are cylindrical. Cutting the strips which are desired in the novel tomography scanner is time-consuming and labor-intensive process, connected with large material waste.

Because of that, cuboidal form is better for the purpose. In the new furnace (Fig. 36), it can be placed horizontally to minimize the pressure of the liquid and monomer pairs (right panel of Fig. 35).

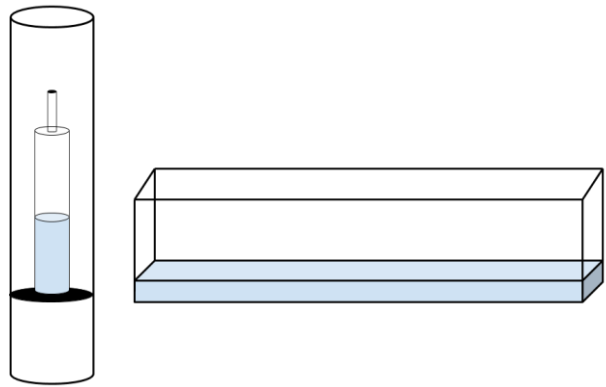

**Figure 35 Left panel: the scheme of arrangement of glass ampoule in pipe furnace. Right panel: the scheme of arrangement of the form in DCF420/spec furnace. Figure is not to scale, the volume of scintillator is marked in blue.**

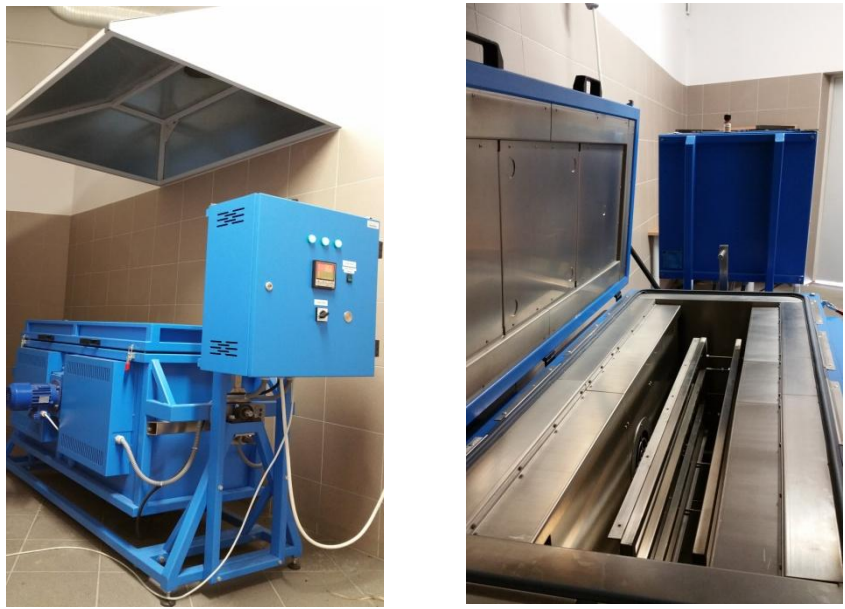

**Figure 36 Photograph of new furnace (left panel) and its interior (right panel).**

The furnace chamber (right panel of Fig. 36) ensures uniform temperature distribution in a whole volume. Precise temperature control (1 °C) and setting whole temperature cycle are possible as well.

The polymerization of scintillating mixture occuring in the furnace leads to obtaining scintillating material in the form of strip. The photograph of the strip in UV light is presented in Fig. 37. The scintillating material is optically homogeneous and does not contain any defects.

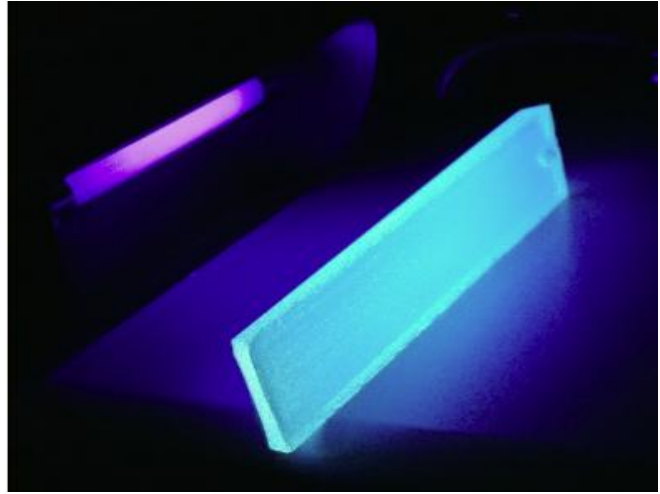

**Figure 37 J-PET scintillator strip exposed to UV light.**

Regarding the development of long J-PET scintillator strips, it is necessary to study the absorption length in the novel scintillating material. This was discussed in chapter 7.1. Optical properties of J-PET scintillator. Attenuation length in the scintillating material can be determined analyzing dependence of light output as a function of position along the strip. Charge of signals at the Compton edge were determined irradiating J-PET and BC-420 [12] scintillators in several points using method described in chapter 7.2. The dependence for 18 cm long J-PET scintillator is plotted in Fig. 38.

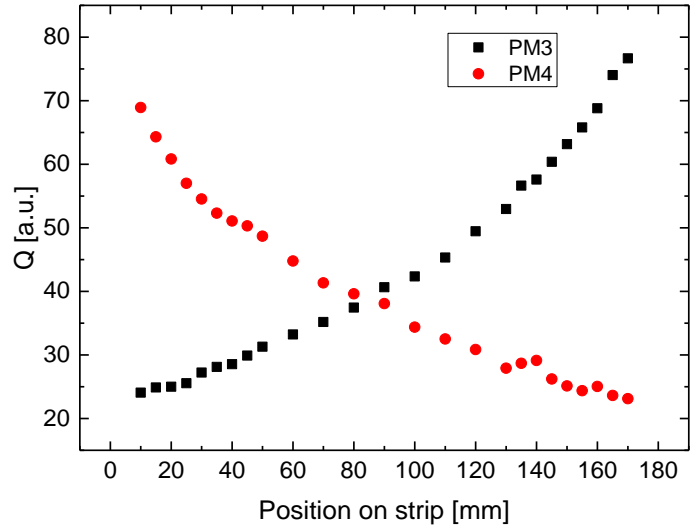

**Figure 38 Charge (Q) of signals at Compton edge for irradiation positions along the J-PET scintillator strip. Charge was calculated for signals registered by photomultipliers at both ends of scintillator (PM3 and PM4).**

Attenuation length can be determined by fitting function given by formula 11 [105] to experimental points:

$$Q = c_1 * \exp(-x/\lambda_1) + c_2 * \exp(-x/\lambda_2) \quad (11).$$

Fitting to experimental points determined for PM 4 is shown in Fig. 39.

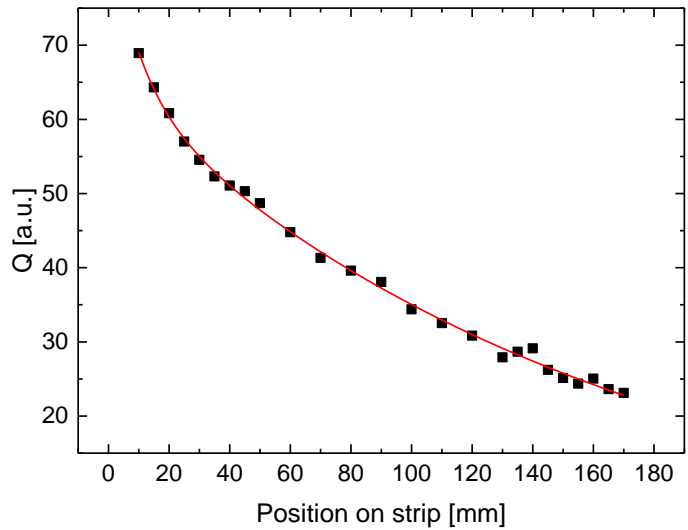

**Figure 39 Function given by formula 11 fitted to experimental points (PM4).**

Such shape of fitted curves is justified by appearance of two components of light attenuation: long ($\lambda_1$) and short ($\lambda_2$), that can be pointed out from fitting. In short scintillators, about 20 cm long, the short component dominates. Attenuation length is dependent on the wavelength [9]. The obtained results refer to blue light, to which vacuum photomultipliers are sensitive.

Function given by formula 11 was fitted also to experimental points of 25 cm long BC-420 scintillator, and parameters $\lambda_1$ and $\lambda_2$ were determined as well. In Tab. 11 absolute values of attenuation parameters determined for J-PET and BC-420 scintillators with pair of photomultipliers for each strip are presented.

Table 11 Attenuation parameters for J-PET and BC-420 scintillators.

| Scintillator/parameter | $\lambda_1$ [mm] | $\lambda_2$ [mm] |
|---|---|---|
| J-PET | 10.3 ± 2.6 | 26.0 ± 5.7 | 158.4 ± 6.5 | 163.0 ± 3.7 |
| BC-420 | 8.8 ± 2.3 | 31.7 ± 7.8 | 492 ± 60 | 308.6 ± 8.2 |

In case of BC-420, mean values of $\lambda_1$ and $\lambda_2$ are comparable with corresponding parameters of 30 cm scintillators determined and described in [105]. For blue light, values of short component: $\lambda_1$ in both scintillators are consistent within error bars. Therefore, light attenuation in scintillating strips, about 20 cm long, is similar. However, the long constant: $\lambda_2$ differs significantly in both scintillators. In case of BC-420 it is more than two times bigger than in J-PET scintillator. This indicates that attenuation in longer J-PET scintillators will be larger in comparison to BC-420.

The described test was carried out for blue light detected by vacuum photomultipliers. Here it is important to note that J-PET is emitting light also in the range between about 330 nm and 370 nm (see Fig. 16), where the attenuation is strong and where BC-420 has no contribution. This is the region of relatively high quantum efficiency of vacuum photomultipliers but not silicon ones. The experiment does not describe attenuation of light of the targeted wavelength, but shorter ones.

Moreover, J-PET plastic scinitllators are prepared from monomer containing inhibitor of polymerization: 4-tert-butylcatechol, in amount of 50 ppm. It is brown solid, yellowish when dissolved, what may cause slight yellowing of polymer. Even such low amount may affect the light propagation and attenuation length in the scintillating material.

J-PET scintillators will be eventually utilized with silicon photomultipliers, sensitive to light of larger wavelength. Maximum of emission of J-PET scintillators is shifted towards longer wavelengths, for which attenuation is lower. Therefore in long J-PET scintillators strips the attenuation length should be larger than in BC-420 of the same length.

## 10. Summary and perspectives

The aim of the thesis was to develop novel plastic scintillator for use in the hybrid J-PET/MR tomograph. Such scintillator ought to be optimized for the superior timing properties. Its emission spectrum should be compatible with the quantum efficiency of silicon photomultipliers. To achieve this goal, scintillators with the novel compositions were designed and produced. Three substances were synthesized and tested as a potential wavelength shifters. Plastic scintillators containing these dopants as a secondary additives were obtained via bulk polymerization of vinyltoluene. One of the additives, 2-(4-styrylphenyl)benzoxazole proved to be very effective as a scintillating dopant. Scintillators containing the additive were a subject of research described in the thesis.

The use of 2-(4-styrylphenyl)benzoxazole as a plastic scintillator dopant is profitable because of its emission spectrum, which maximum is shifted towards longer wavelengths in comparison to state-of-the-art commercial scintillators. Because of that, J-PET scintillator is better matched to digital silicon photomultipliers spectral characteristics, which will be utilized in J-PET/MR tomograph. Larger emission wavelength provides smaller absorption coefficient. Thus attenuation length of J-PET scintillator is relatively longer, which is desirable considering long scintillator strips.

Further on, more detailed tests and analysis of novel scintillators were conducted. Series of scintillators with 2,5-diphenyloxazole (PPO) as a primary fluor and different concentrations of 2-(4-styrylphenyl)benzoxazole as a wavelength shifter, were prepared. Light output of scintillators were determined, basing on interaction with gamma quanta originating from $^{22}$Na source. Optimal concentration of novel wavelength shifter in plastic scintillator, providing maximal light output is equal to 0.05 wt. ‰ (0.05J-PET scintillator).

Scintillators containing smaller amounts of the WLS exhibit lower light output because inefficient energy transfer, while for those with higher than 0.05 ‰ WLS concentration, the concentration quenching occurs. Such light output dependence on the WLS concentration was observed also in case of commercial wavelength shifter POPOP.

Light signals in the 0.05J-PET scintillator were analyzed as well. Rise and decay times of the signals were determined to be equal to 0.50 ns and 1.91 ns, respectively. Rise time of pulses appearing in the J-PET scintillator is equal to the rise time of signals in BC-420 scintillator, however decay time is longer by about 0.4 ns. BC-420 is one of the best scintillators by Saint Gobain considering time properties. When comparing J-PET to other scintillators produced by the company, its decay time has the typical value.

Since small molecular weight of scintillators polymeric matrix decreases its light output, it was essential to determine molecular weight of J-PET scintillator. It was established that molecular weight of J-PET scintillator, exceeding $10^5$ u, has sufficiently high value for which the light output value is not being affected anymore. Considering polymer influence on a scintillator as a system, light output is maximal.

Structure of J-PET scintillator was studied using two methods: Positron Annihilation Lifetime Spectroscopy (PALS) and Differential Scanning Calorimetry (DSC). Glass transition temperature ($T_g$), temperature in which structural transition occur ($T_\gamma$) and softening point which is taken as a maximal temperature at which scintillator can act was determined. In order to discuss the influence of scintillating dopants on the structure, results obtained for J-PET scintillator and pure polymeric matrix samples were compared. Significant discrepancy between temperatures determined by PALS and DSC were observed. The differences are related to the chosen experimental techniques and their limitations.

As construction of the J-PET/MR hybrid tomograph requires hundredths of at least 50 cm long plastic scintillators, the next step of conducted research was setting the conditions for larger than studied so far scintillating samples production. Design and preparation of the special reactor was necessary. Many difficulties were connected with the choice of proper material for the form. The furnace for polymerization has been changed because the geometry of the previous one prevented the preparation of large scintillator strips. Apart from purely geometrical issues regarding the form and furnace construction one has to take into account the fact that the chemical composition, especially wavelength shifter concentration in scintillating material is not a fixed value in terms of size and shape of synthesized sample [31]. Therefore, optimization of the WLS concentration has to be performed each time the geometry of the scintillator is to be changed.

One of the factors, which should be determined for long J-PET scintillator strips, is attenuation length. As the novel J-PET scintillator has different spectral characteristics of emitted scintillation light, that is shifted towards longer wavelengths than for commercial products, also the setup for the wavelength dependent light attenuation measurement should be modified. So far used vacuum photomultipliers are less sensitive for the longer wavelengths, whereas silicon photomultipliers quantum efficiency reaches maximum in that region. Therefore, a next step for better attenuation length determination for J-PET scintillator will be silicon photomultipliers use in the experimental system.

Properties of J-PET scintillator, like light output, rise and decay time, emission spectrum and H:C ratio were compared to properties of commercially available scintillators produced by Saint Gobain in Tab. 12. No significant difference in any of the value was observed. This indicates that properties of J-PET scintillator are similar to properties of commercial plastic scintillators. Emission spectrum, which is specific for particular scintillator, is important in the view of scintillator application, because it has to be matched to the particular scintillation light detector which is used in the experiment. H:C ratio is typical for plastic scintillators.

**Table 12 Properties of several scintillators of Saint Gobain: BC-420, BC-404, BC-408 and the J-PET scintillator [12].**

| Properties | BC-420 | BC-404 | BC-408 | J-PET |
|---|---|---|---|---|
| Light output [% of Anthracene] | 64 | 68 | 64 | 64 |
| Rise time [ns] | 0.5 | 0.7 | 0.9 | 0.5 |
| Decay time [ns] | 1.5 | 1.8 | 2.1 | 1.9 |
| Maximum of emission wavelength [nm] | 391 | 408 | 425 | 404 |
| H:C ratio | 1.102 | 1.107 | 1.104 | 1.104 |

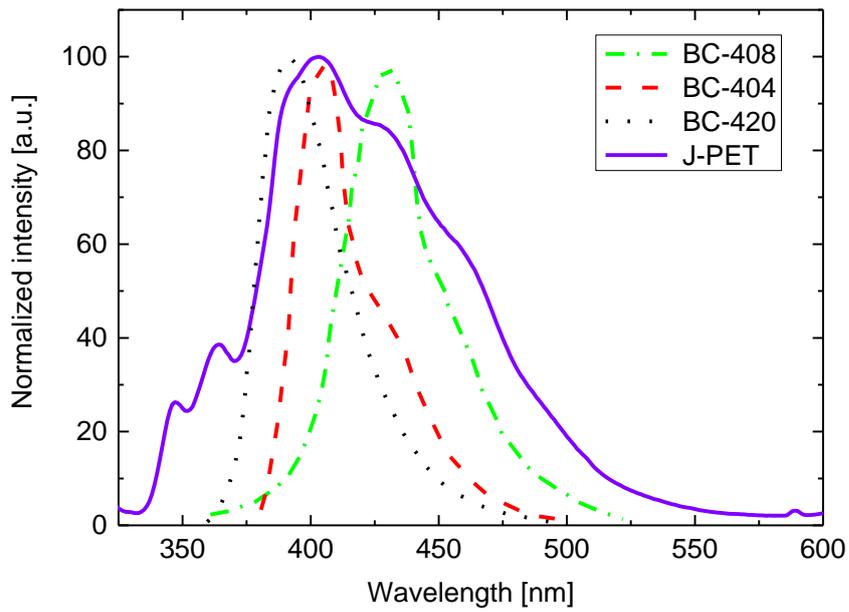

**Figure 40 Emission spectra of BC-408 (green dashed - dot line), BC-404 (red dashed line), BC-420 (black dotted line) and J-PET scintillator (violet solid line).**

Regarding the emission spectra of the considered scintillators, the best matching to the quantum efficiency wavelength dependency of silicon photomultipliers is achieved for BC-408 scintillator. According to Fig. 15, the highest quantum efficiency is for around 450 nm. However, having in mind further application in the detectors relying on the signals time measurements, one has to carefully look also at the timing properties of the scintilators, for which the BC-408 is the worst within the group being compared.

Cooperation with worldwide companies, like Saint Gobain [12] or Eljen Technology [13] indicates that even up to 40 % of purchased scintillator strips contained optical inhomogeneities visible by the eye or when exposed to UV light. Such defects disturb isotropic propagation of light in scintillator material and significantly decrease the light output. Therefore, less amount of scintillation light reaches photoelectric converters what makes the whole detector less efficient for the radiation and particles detection.

Scintillators containing defects need to be changed for the homogeneous ones, however the procedure itself takes time. Moreover such method of scintillators preparation entails a large material waste.

Therefore, the process should be optimized considering assortment of the reactor as well as condition of polymerization to obtain a large fraction of good quality scintillators minimizing material waste. That also decreases costs of final product, ready for application.

Summarizing, with the presented dissertation it has been proven that in the laboratory conditions it is possible to develop plastic scintillators characterized by the parameters fulfilling the conditions for further application in J-PET/MR tomography as well as detectors for particle physics research.

# List of abbreviations

PET - Positron Emission Tomography

MRI - Magnetic Resonance Imaging

LOR - Line of Response

BGO - $BiGe_3O_{12}$

LSO:Ce - $Lu_2SiO_5:Ce^{3+}$

LYSO:Ce - $(Lu,Y)_2SiO_5:Ce^{3+}$

GSO - $Gd_2SiO_5:Ce^{3+}$

PM - photomultiplier

CT - Computed Tomography

SiPM - silicon photomultiplier

WLS - wavelength shifter

PVT - polyvinyltoluene

PS - polystyrene

PTP - p-terphenyl

PPO - 2,5 -diphenyloxazole

PPD - 2,5-diphenyl-1,3,4-oxadiazole

BBD - 2,5-bis(4-biphenyl)-1,3,4-oxadiazol

PBD - 2-phenyl-5(4-biphenyl)-1,3,4-oxadiazole

BPBD - 2-(4-tert-buthylphenyl)-5-(4-biphenylo)-1,3,4-oxazdiazole

POPOP - 1,4-bis(5-phenyl-2-oxazolyl)benzen

DM-POPOP - 1,4-bis(4-methyl-5-phenyl-2-oxazolyl)benzene

Bis-MSB - 1,4-bis(2-methylostyryl)benzene

BBO - 2,5-di(4-biphenylo)oxazole

DPS - trans-4,4'-diphenylstilbene

DPA - 9,10-diphenylanthracene

PALS - Positron Annihilation Lifetime Spectroscopy

DSC - Differential Scanning Calorimetry

DMAPOP - 2-(4'-*N,N*-dimethylaminophenyl)oxazolo[4,5-*b*]pirydyne

PPA - polyphosphoric acid

TLC - Thin Layer Chromatography

P(OEt)$_3$ - triethylphosphine

TTS - Transit Time Spread

T$_g$ - glass transition temperature

p-Ps - para-positronium

o-Ps - ortho-positronium

$\tau_3$ - o-Ps lifetime

I$_3$ - o-Ps intensity

# Acknowledgements

I would like to express my special appreciation and thanks to prof. Paweł Moskal and dr Andrzej Kochanowski for supervising the thesis, valuable advices and all opportunities I was given to carry out my research and my dissertation.

I would like to express my deepest gratitude to Scientific Council and all Professors of Institute of Metallurgy and Materials Science, Polish Academy of Sciences in Kraków, for the opportunity of realization Ph. D. studies and all kindness I have experienced.

I warmly thank members of J-PET collaboration for their great help in hundreds of hours of measurements, data analysis, scientific discussions and nice atmosphere during everyday meetings. Especially I am grateful to mgr Szymon Niedźwiecki for preparation of software for charge spectra analysis, to mgr Kamil Dulski for calculations of parameters describing timing properties of scintillators, to mgr Neha Gupta Sharma for calculating average signals appearing in scintillators. Special thanks go to dr Bartosz Głowacz for organizing an access to the chemical laboratory of Atomic Optics Department, cooperation in designing reactors and in scintillators strips development, and to dr inż. Marcin Zieliński for help with conceptual work and preparation of scientific projects. I also thank mr Wojciech Migdał for mechanical machining of scintillators.

I am sincerely grateful to dr hab. Monika Marzec and mgr Jakub Fitas for the opportunity of conducting DSC measurements and help with them.

I would also like to thank Members of Research Group from Maria Skłodowska - Curie University in Lublin, especially prof. Bożena Jasińska and dr Bożena Zgardzińska for the opportunity of conducting PALS measurements, wonderful cooperation, valuable comments and suggestions.

I gratefully acknowledge dr hab. Andrzej Danel and the research group for cooperation on a new scintillation additive.

Words cannot express how grateful I am to my dear Friends and Colleagues who I have known from years and who I have met during my Ph. D. studies. Thank you for Your great support and making last four years an excellent time.

And finally, my sincere thanks go to my Parents for Their endless love and encouragement. I would like to thank Michał for the supporting serenity. Also I thank Julia and Natalia for positive energy.